\newcommand{\diracslash}[1]{#1\llap{/\kern2pt}}

\newcommand{\be}{\begin{equation}}
\newcommand{\ee}{\end{equation}}
\newcommand{\bea}{\begin{eqnarray}}
\newcommand{\eea}{\end{eqnarray}}
\newcommand{\ba}[1]{\begin{array}{#1}}
\newcommand{\ea}{\end{array}}

\newcommand{\bt}{\begin{tabular}}
\newcommand{\et}{\end{tabular}}

\newcommand{\beas}{\begin{eqnarray*}}
\newcommand{\eeas}{\end{eqnarray*}}

\documentclass[preprint,prd,aps,floats,nofootinbib,floatfix]{revtex4}

\DeclareSymbolFont{rsfs}{U}{rsfs}{m}{n}
\DeclareSymbolFontAlphabet{\mathrsfs}{rsfs}
\usepackage{amsmath}
\usepackage{slashed}
\usepackage{graphicx}
\usepackage{cleveref}
\usepackage{epstopdf}
\usepackage{color}
\usepackage{ulem}

\begin{document}

\title{Octet Baryon Masses and Magnetic Moments in Hot and Dense Isospin Asymmetric Nuclear Matter}

 \author{Harpreet Singh}
\email{harpreetmscdav@gmail.com}
\affiliation{Department of Physics, Dr. B R Ambedkar National Institute of Technology Jalandhar, 
 Jalandhar -- 144011, Punjab, India}
 \author{Arvind Kumar}
\email{iitd.arvind@gmail.com, kumara@nitj.ac.in}
\affiliation{Department of Physics, Dr. B R Ambedkar National Institute of Technology Jalandhar, 
 Jalandhar -- 144011, Punjab, India}
  \author{Harleen Dahiya}
\email{dahiyah@nitj.ac.in}
\affiliation{Department of Physics, Dr. B R Ambedkar National Institute of Technology Jalandhar, 
 Jalandhar -- 144011, Punjab, India}

\def\be{\begin{equation}}
\def\ee{\end{equation}}
\def\bearr{\begin{eqnarray}}
\def\eearr{\end{eqnarray}}
\def\zbf#1{{\bf {#1}}}
\def\bfm#1{\mbox{\boldmath $#1$}}
\def\hf{\frac{1}{2}}
\def\kp{\zbf k+\frac{\zbf q}{2}}
\def\km{-\zbf k+\frac{\zbf q}{2}}
\def\hwo{\hat\omega_1}
\def\hwt{\hat\omega_2}

\begin{abstract}

 Medium modification of the magnetic moments of octet baryons in the isospin asymmetric nuclear medium at finite temperature has been calculated using medium modified quark and baryon masses derived in chiral SU(3) quark mean field model. The magnetic moments of baryons are found to vary significantly as a function of density of nuclear medium as well as with the increase in isospin asymmetry of the medium. The rise of temperature is found to decrease the effect of isospin asymmetry on masses and magnetic moments of baryons. The results of present investigation may be helpful to understand the experimentally observed values of octet baryon magnetic moments at different experimental facilities.


\end{abstract}

\maketitle

\section{Introduction}
Nuclear matter can be defined as a hypothetical system composed of nucleons interacting without Coulomb forces. 
A fixed ratio of protons and neutrons can be idealized for studying the nuclear matter \cite{bordbar,horst,friedman}. For example, in case of symmetric nuclear matter studies, the ratio of number of protons and neutrons is taken to be $1:1$. The study of properties of nuclear matter as a function of the ratio of number of neutrons to the number of protons (which leads to isospin asymmetry), density and temperature of the nuclear medium has been a fundamental problem for theoretical as well as experimental studies \cite{horst}.  
Many problems in Nuclear Physics such as nucleon-nucleon interactions in dense asymmetrical nuclear matter, neutron star mass and radius, neutron skin of nuclear matter, stability of asymmetric nuclei etc., require calculations for isospin asymmetric nuclear matter \cite{bordbar,horst}. The astrophysical studies in particular have a close relationship with the investigation of isospin asymmetric nuclear matter properties \cite{friedman}. The dense nuclear matter under the influence of external magnetic field of the order of $10^{18}$ G, (present in magnetars) can also lead to isospin splitting between baryons, which will also modify the static properties of baryons in these systems \cite{ryu1,mcimcqcd}.

The advent of radioactive beam facilities around the world, such as FAIR at GSI Germany \cite{friese}, Cooling Storage Ring (CSR) at HIRFL in China \cite{zhan}, Radioactive Ion Beam (RIB) Factory at RIKEN in Japan \cite{yano}, SPIRAL2/GANIL in France and Facility for Rare Isotope Beams (FRIB) in United States, is a step forward for exploring properties of isospin asymmetric nuclear matter under extreme conditions of large isospin asymmetry. 
The heavy-ion collision experiment at RHIC and LHC explore the region of QCD phase diagram at low baryonic density and high temperature of the medium, whereas CBM experiment of FAIR project \cite{friese} at GSI may explore the region of high baryonic density and moderate temperatures \cite{rahul}. 

The thermodynamical properties of asymmetric nuclear matter such as Equation of State (EoS) and the symmetry energy have been widely studied \cite{eng,dave,sahu,lopez,shlomo,fabio,lenske,jeremy} (also see references their in). Beside the study of thermodynamical properties of asymmetric nuclear matter, the dependence of structural properties of baryons on isospin asymmetry of the nuclear medium plays a major role in baryonic studies. Magnetic moments of octet baryons is one of the major structural property which is directly related to the internal structure of baryons. It plays a crucial role in understanding of hadronic structure from underlying theory of strong interactions, QCD (Quantum Chromo Dynamics) and also in context of phenomenological models \cite{savage}. The study of magnetic moments of octet baryons has a long standing history. In free space, the magnetic moments of octet baryons have been studied by various authors using different theoretical models \cite{felix,wrb,hack,jg,jun,lks,ss}. However, the idea of medium dependence of magnetic moments of baryons has recently gained interest. For example in Ref. \cite{mulders} the magnetic moment of proton in $^{12}\text{C}$ seemed to be enhanced by about $25\%$ in nuclear medium as compared to its value in free space.  
 
In order to study the properties of baryons in dense asymmetric nuclear matter, it is natural to start with a model which uses quarks as fundamental degrees of freedom \cite{lawley}. Many models on these lines have been constructed such as quark meson coupling model \cite{guic1,guic2,saito,panda}, chiral soliton model \cite{smith}, cloudy bag model \cite{mil}, Nambu-Jona-Lasinio (NJL) model \cite{bub}, chiral SU(3) quark mean field (CQMF) \cite{wang,wang2} model etc. Magnetic moments of octet baryons at finite temperature and density of nuclear medium have been studied using the quark meson coupling model and the modified quark meson coupling model \cite{ryu1,ryu2}. In our recent work \cite{happy}, we have calculated the magnetic moments of octet baryons at finite temperature and density of symmetric nuclear matter using chiral SU(3) quark mean field model. Results obtained were comparable to those available in literature and the magnitudes of magnetic moments of baryons were found to increase as a function of density in the symmetric nuclear matter at zero as well as at finite temperatures.
 
In the present work, we have used chiral SU(3) quark mean field model \cite{wang,wang2} to study medium modification of magnetic moments of octet baryons in isospin asymmetric nuclear matter at finite temperature and different densities of the medium. Mean field approximation method is known to be  thermodynamically consistent because it satisfies the relevant thermodynamic identities and viral theorem \cite{serot2} on the one hand whereas, on the other hand, one can easily determine the model parameters analytically from the specified set of values of nuclear matter properties at zero temperature \cite{bodmer,serot2}, making it easy to study variations of the model and model dependent properties of baryons \cite{horst}. 
Mean field models enable one to study the properties of baryons in terms of quark and gluon degrees of freedom \cite{wang,wang2,asywang}, which are of central importance in recent studies and in the present work as well.   

Since the isovector channel is very important for the study of asymmetric nuclear matter, we have included the contribution from the scalar isovector $\delta$ meson field in chiral SU(3) quark mean field model. The properties like isospin diffusion in heavy ion collisions \cite{chen} and size of neutron skins in finite nuclei \cite{typel} are  also influenced by the isovector part of nuclear medium.  
In the mean field studies, the inclusion of scalar isovector $\delta$ meson field and its influence on low density asymmetric matter was studied \cite{liu,greco,kubis}. The $\delta$ meson is found to play an important role in strongly asymmetric nuclear matter at higher densities. 
In Relativistic Mean field (RMF) theory, the inclusion of $\delta$ field is found to alter the threshold characteristics of mixed phase \cite{grigor1}, leading to non-negligible changes in the  quark first order transitions \cite{grigor2}. 
It will be interesting to observe the behavior of $\delta$ field in chiral SU(3) quark mean field model \cite{asywang}. 

 In our scheme of study, we will calculate octet baryon masses and  magnetic moments using the medium modified values of masses of constituent quarks in the asymmetric nuclear matter at finite temperature and density. We will also include the contribution to the total effective magnetic moment of baryons from sea quarks and orbital angular momentum of sea quarks along with the contribution from valence quarks, which have been successfully used to calculate octet baryon magnetic moments in free space \cite{cheng1,har} as well as in symmetric nuclear matter \cite{happy}. 

           The outline of the paper is as follows : In \cref{masscalcua} we will apply CQMF model along with the contribution from $\delta$ meson field to find the effective quark masses at finite temperature and density of asymmetric nuclear medium, and hence, calculate effective baryon octet masses. We will discuss the effect of valence quark  polarization, quark sea polarization and quark sea orbital angular momentum on the magnetic moments of baryons in \cref{masscalcub}. Section \ref{results} is devoted to numerical calculations and results. Section \ref{summ} includes the summary of present work.
     
\section{Model} 
\subsection{CQMF for asymmetric nuclear matter} \label{masscalcua}
We consider the chiral SU(3) quark mean field model to study isospin asymmetric nuclear matter which uses the quarks and mesons as basic degrees of freedom. The quarks are assumed to be confined in the baryons by an effective potential. Quark meson interactions as well as meson self interactions are based on the chiral SU(3) symmetry. The constituent quarks and mesons (except pseudoscalar mesons) obtain their masses through the phenomenon of spontaneous symmetry breaking. The pseudoscalar mesons get their masses through the introduction of an explicit symmetry breaking term in meson self interaction which satisfies partially conserved axial-vector current relations \cite{happy,wang}.

To study the structure of baryons in chiral limit and to explore it in quark degrees of freedom, the quarks are divided into two parts, left-handed quarks `$q_L$' and right-handed quarks `$q_R$' : $q=q_L+q_R$. They transform under global $\text{SU(3)}_L \times \text{SU(3)}_R$ transformations $`L$' and $`R$' as 
\begin{equation}
q_{L} \rightarrow q_{L}^{\prime }\,=\,L\,q_{L},~~~~~
q_{R} \rightarrow q_{R}^{\prime }\,=\,R\,q_{R},\,
\end{equation} 
where `$L$' and `$R$' are given as 
\begin{equation}
L(\alpha_L)=\text{exp}\left[ i\sum {\alpha}_L^a {\lambda}_{La}\right] ,\,~~~~~~~~R(\alpha_R)=\text{exp}\left[ i\sum {\alpha}_R^b {\lambda}_{Rb}\right] ,
\end{equation}
$\alpha_L$ and $\alpha_R$ represent space-time independent parameters and $\lambda_L$ and $\lambda_R$ are Gell-Mann matrices and are written as
\begin{equation}
\lambda_L=\lambda\frac{( 1-\gamma_5)}{2}, \,~~~~~~~~
\lambda_R=\lambda\frac{( 1+\gamma_5)}{2}.
\end{equation}
The nonents of spin-0 scalar ($\Sigma$) and pseudoscalar ($\Pi$) mesons can be written in compact form by using Gell-Mann matrices as
\begin{equation}
M(M^{\dagger})=\Sigma \pm i\Pi =\frac{1}{\sqrt{2}}\sum_{a=0}^{8}
\left( s_{a}\pm i p _{a}\right) \lambda _{a},
\end{equation}
where $\lambda _{a}$ ($a=1,......,8$) are the Gell-Mann matrices with $\lambda _{0}=\sqrt{\frac{2}{3}}I$, $s_{a}$ and $p_{a}$ are the nonets of scalar and pseudoscalar
mesons, respectively. The plus and minus signs stand for  $M$ and $M^{\dagger}$ respectively. $M$ and $M^{\dagger}$ transform under chiral $\text{SU(3)}$ transformation as
\begin{equation}
M\rightarrow M^{\prime }=LMR^{\dagger},
\end{equation}
\begin{equation}
 M^{\dagger}\rightarrow
M^{{\dagger}^{\prime }}=RM^{\dagger}L^{\dagger}.
\end{equation}
In the similar way, spin-1 mesons are defined by 
\begin{equation}
l_{\mu }(r_{\mu })=\frac{1}{2}\left( V_{\mu }\pm A_{\mu }\right)
= \frac{1}{2\sqrt{2}}\sum_{a=0}^{8}\left( v_{\mu}^{a}\pm a^{a}_{\mu}
\right) \lambda_{a},
\end{equation}
where $v_{\mu}^a$ and $a_{\mu}^a$ are nonets of vector and pseudovector mesons respectively.
The alternative plus and minus signs are for $l_{\mu}$ and $r_{\mu}$ which transform under chiral $\text{SU(3)}$ transformation as 
\begin{equation}
l_{\mu}\rightarrow l_{\mu }^{\prime }=Ll_{\mu }L^{\dagger},
\end{equation}
\begin{equation}
r_{\mu}\rightarrow r_{\mu }^{\prime }=Rr_{\mu }R^{\dagger}.
\end{equation}
 The physical states for scalar and vector mesons are explicitly represented as
\begin{equation}
\Sigma = \frac1{\sqrt{2}}\sum_{a=0}^8 s_a \, \lambda_a=\left(
\begin{array}{lcr}
\frac1{\sqrt{2}}\left(\sigma+\delta^0\right) & \delta^{+} & \kappa^{*+} \\
\delta^- & \frac1{\sqrt{2}}\left(\sigma-\delta^0\right) & \kappa^{*0} \\
\kappa^{*-} & \bar{\kappa}^{*0} & \zeta
\end{array}
\right),
\end{equation}
and
\begin{equation}
V_\mu = \frac1{\sqrt{2}}\sum_{a=0}^8 v_\mu^a \, \lambda_a=\left(
\begin{array}{lcr}
\frac1{\sqrt{2}}\left(\omega_\mu+\rho_\mu^0\right)
& \rho_\mu^+ & K_\mu^{*+}\\
\rho_\mu^- & \frac1{\sqrt{2}}\left(\omega_\mu-\rho_\mu^0\right)
& K_\mu^{*0}\\
K_\mu^{*-} & \bar{K}_\mu^{*0} & \phi_\mu
\end{array}
\right).
\end{equation}
In the similar manner, we can write pseudoscalar nonet ($\Pi$) and pseudovector nonet ($A_{\mu}$). 
The total effective Lagrangian density in chiral $\text{SU(3)}$ quark mean field model is written as 
\begin{equation}
{\cal L}_{{\rm eff}} \, = \, {\cal L}_{q0} \, + \, {\cal L}_{qm}
\, + \,
{\cal L}_{\Sigma\Sigma} \,+\, {\cal L}_{VV} \,+\, {\cal L}_{\chi SB}\,
+ \, {\cal L}_{\Delta m} \, + \, {\cal L}_{c}, \label{totallag}
\end{equation}
where ${\cal L}_{q0} =\bar q \, i\gamma^\mu \partial_\mu \, q$ represents the free part of massless quarks, ${\cal L}_{qm}$ is the chiral SU(3)-invariant quark-meson interaction term and is written as \cite{wang,wang2,happy}
\begin{align}
{\cal L}_{qm}=g_s\left(\bar{\Psi}_LM\Psi_R+\bar{\Psi}_RM^{\dagger}\Psi_L\right)
-g_v\left(\bar{\Psi}_L\gamma^\mu l_\mu\Psi_L+\bar{\Psi}_R\gamma^\mu
r_\mu\Psi_R\right)~~~~~~~~~~~~~~~~~~~~~~~  \nonumber \\
=\frac{g_s}{\sqrt{2}}\bar{\Psi}\left(\sum_{a=0}^8 s_a\lambda_a
+ i \gamma^5 \sum_{a=0}^8 p_a\lambda_a
\right)\Psi -\frac{g_v}{2\sqrt{2}}
\bar{\Psi}\left( \gamma^\mu \sum_{a=0}^8
 v_\mu^a\lambda_a
- \gamma^\mu\gamma^5 \sum_{a=0}^8
a_\mu^a\lambda_a\right)\Psi, \label{quarkmesons}
\end{align}
where $\Psi=\left(\begin{array}{lcr}
u
& \\
d
& \\
s
\end{array}\right)
$. 
The various coupling constants used in the present work are
\begin{align}
\frac{g_s}{\sqrt{2}}
= &g_{\delta}^u = -g_{\delta}^d = g_\sigma^u = g_\sigma^d  =
\frac{1}{\sqrt{2}}g_\zeta^s, \label{relation}
~~~~~g_{\delta}^s = g_\sigma^s = g_\zeta^u = g_\zeta^d = 0 \, ,\\
&\frac{g_v}{2\sqrt{2}}
= g_{\rho}^u = -g_{\rho}^d = g_\omega^u = g_\omega^d,
~~~~~~~g_\omega^s = g_{\rho}^s  = 0 .
\end{align}
These parameters are calculated by fitting energy of nuclear matter at saturation density ($\rho_0=0.16$ $\rm{fm^{-3}}$).
The chiral-invariant scalar and vector meson self interaction terms ${\cal L}_{\Sigma\Sigma}$ and ${\cal L}_{VV}$, within mean field approximation \cite{wang} and including scalar isovector meson $\delta$ are written as
\begin{align}
{\cal L}_{\Sigma\Sigma} =& -\frac{1}{2} \, k_0\chi^2
\left(\sigma^2+\zeta^2+\delta^2\right)+k_1 \left(\sigma^2+\zeta^2+\delta^2\right)^2
+k_2\left(\frac{\sigma^4}{2} +\frac{\delta^4}{2}+3\sigma^2\delta^2+\zeta^4\right)\nonumber \\ 
&+k_3\chi\left(\sigma^2-\delta^2\right)\zeta 
 -k_4\chi^4-\frac14\chi^4 {\rm ln}\frac{\chi^4}{\chi_0^4} +
\frac{\xi}
3\chi^4 {\rm ln}\left(\left(\frac{\left(\sigma^2-\delta^2\right)\zeta}{\sigma_0^2\zeta_0}\right)\left(\frac{\chi^3}{\chi_0^3}\right)\right), \label{scalar0}
\end{align}    
and
\begin{equation}
{\cal L}_{VV}=\frac{1}{2} \, \frac{\chi^2}{\chi_0^2} \left(
m_\omega^2\omega^2+m_\rho^2\rho^2\right)+g_4\left(\omega^4+6\omega^2\rho^2+\rho^4\right), \label{vector}
\end{equation}
respectively.
The parameter $\xi$ originates from logarithmic term used in scalar meson self interaction Lagrangian density and can be obtained using QCD $\beta$-function at one loop level for three colors and three flavors \cite{papag}. 
 The constants $k_0, k_1, k_2, k_3$ and $k_4$ appearing in \cref{scalar0} are respectively determined using $\pi$ meson mass ($m_{\pi}$), $K$ meson mass ($m_K$) and the average mass of $\eta$ and $\eta^{'}$ mesons. 
 
The vacuum expectation values of scalar meson fields $\sigma$ and $\zeta$, i.e., $\sigma_0$ and $\zeta_0$ are constrained because of spontaneous breaking of chiral symmetry and are represented as 
\begin{align}
\sigma_0= -f_{\pi} ~~{\rm and}~~~~  \zeta_0= \frac{1}{\sqrt{2}}\left( f_{\pi}-2f_{K}\right), 
\end{align} 
where $f_{\pi}$ and $f_K$ are pion and kaon leptonic decay constants, respectively.
In our calculations we have chosen $f_{\pi}=92.8$ MeV and $f_{K}=115$ MeV with the corresponding values of $\sigma_0$ and $\zeta_0$ being $-93.49$ MeV and $-97.98$ MeV  respectively. 
 The vacuum value of dilaton field $\chi_0=254.6$ MeV and the coupling constant $g_4=37.4$, are chosen so as to fit reasonable effective nucleon mass.
 
 The Lagrangian density ${\cal L}_{\chi SB}$ in \cref{totallag} is the explicit symmetry breaking term, which is introduced to incorporate non-vanishing pesudoscalar meson masses and it satisfies the partial conserved axial-vector current relations for $\pi$ and $K$ mesons \cite{wang}. It is expressed as
\begin{equation}\label{L_SB}
{\cal L}_{\chi SB}=\frac{\chi^2}{\chi_0^2}\left[m_\pi^2f_\pi\sigma +
\left(
\sqrt{2} \, m_K^2f_K-\frac{m_\pi^2}{\sqrt{2}} f_\pi\right)\zeta\right].
\end{equation}
Vacuum masses of `$u$', `$d$' and `$s$' quarks are generated by the vacuum expectation values of $\sigma$ and $\zeta$ mesons scalar fields. In order to find constituent strange quark mass correctly, an additional mass term $\Delta m$, 
which explicitly breaks the chiral symmetry, is included in \cref{totallag} 
through Lagrangian density \cite{wang,happy}
\begin{equation}
{\cal L}_{\Delta m} = - (\Delta m) \bar \psi S_1 \psi,
\end{equation}
where the strange quark matrix operator $S_1$ is defined as
\begin{equation}
S_1 \, = \, \frac{1}{3} \, \left(I - \lambda_8\sqrt{3}\right) =
{\rm diag}(0,0,1).
\end{equation}
Thus, the relations for vacuum masses of quarks are 
\begin{align}
\label{qvacmass}
m_u=m_d=-g_{\sigma}^q \sigma_0=-\frac{g_s}{\sqrt{2}}\sigma_0,
\hspace*{.5cm} \mbox{and} \hspace*{.5cm}
m_s=-g_{\zeta}^s \zeta_0 + \Delta m. 
\end{align}
The values of coupling constant $g_s$ and additional mass term $\Delta m$ in \cref{qvacmass} can be calculated by taking $m_u=m_d=313$ MeV and $m_s=489$ MeV as the vacuum masses of quarks. 
The constituent quarks of baryons are confined in baryons by confining scalar-vector potential $\chi_c$. The corresponding Lagrangian density is written as 
\begin{align}
{\cal L}_{c} = -  \bar \psi \chi_c \psi,
\end{align}
where the scalar-vector potential $\chi_c$ is given by \cite{wang}  
\begin{align}
\chi_{c}(r)=\frac14 k_{c} \, r^2(1+\gamma^0) \,.   \label{potential}
\end{align}
The coupling constant $k_c$ is taken to be $100 \, \text{MeV}. \text{fm}^{-2}$.

In order to investigate the properties of asymmetric nuclear matter at finite temperature and density, we will use mean field approximation \cite{wang}. The Dirac equation, under the influence of meson mean field, for the quark field $\Psi_{qi}$ is given as 
\begin{equation}
\left[-i\vec{\alpha}\cdot\vec{\nabla}+\chi_c(r)+\beta m_q^*\right]
\Psi_{qi}=e_q^*\Psi_{qi}, \label{Dirac}
\end{equation}
where the subscripts $q$ and $i$ denote the quark $q$ ($q=u, d, s$)
in a baryon of type $i$ ($i=N, \Lambda, \Sigma, \Xi$)\,
and $\vec{\alpha}$\,, $\beta$\, are usual Dirac matrices.
The effective quark mass $m_{q}^*$ is defined as
\begin{equation}
m_q^*=-g_\sigma^q\sigma - g_\zeta^q\zeta - g_\delta^q\delta + m_{q0}, \label{qmass}
\end{equation}
where $m_{q0}$ is zero for non-strange `$u$' and `$d$' quarks, whereas for strange `$s$' quark $m_{q0}=\Delta m=29$ MeV. Effective energy of particular quark under the influence of meson field is given as,
$e_q^*=e_q-g_\omega^q\omega-g_\rho^q\rho \,$ \cite{wang}. For the  confining potential defined by \cref{potential}, the analytical expression for effective energy of quark $e_q^*$ will be
\begin{align}
e_q^*=m_q^*+\frac{3\sqrt{k_c}}{\sqrt{2(e_q^*+m_q^*)}}. \label{qenergy}
\end{align} 
The effective mass of baryons can be calculated using effective quark masses $m_q^*$, using the relation  
\begin{align}
M_i^*=\sqrt{E_i^{*2}- <p_{i \, \text{cm}}^{*2}>}\,, \label{baryonmass}
\end{align} 
where
\begin{equation} 
E_i^*=\sum_qn_{qi}e_q^*+E_{i \, \text{spin}},    \label{energy}
\end{equation}
 is the effective energy of $i^{th}$ baryon in the nuclear medium. Further, $E_{i \, \text{spin}}$ is the correction to baryon energy due to spin-spin interaction of constituent quarks and takes the following values for different octet baryon multiplets   
\begin{align}
E_{N \, \text{spin}} = - 482 \,\,\, {\rm MeV}\,,
E_{\Lambda \, \text{spin}} = - 756.9 \,\,\, {\rm MeV}\,, 
E_{\Sigma \, \text{spin}} =  - 531 \,\, {\rm MeV}\,, 
E_{\Xi \, \text{spin}} =  - 705 \,\, {\rm MeV}\, \nonumber.
\end{align}
These values are determined to fit the respective vacuum values of baryon masses. 
In \cref{baryonmass}, $<p_{i \, \text{cm}}^{*2}>$ is the spurious center of mass motion \cite{barik1,barik2}.
 To study the equations of motion for mesons at finite temperature and density, we consider the thermodynamic potential as
\begin{equation}
\label{thermo}
\Omega = -\frac{g_N k_{B}T}
{(2\pi)^3} \sum_{N = p\,, n\, }
\int_0^\infty d^3k\biggl\{{\rm ln}
\left( 1+e^{- [ E^{\ast}(k) - \nu_B ]/k_{B}T}\right) \\
+ {\rm ln}\left( 1+e^{- [ E^{\ast}(k)+\nu_B ]/k_{B}T}
\right) \biggr\} -{\cal L}_{M},   
\end{equation}
where
\begin{equation}
{\cal L}_{M} \, = 
{\cal L}_{\Sigma\Sigma} \,+\, {\cal L}_{VV} \,+\, {\cal L}_{\chi SB}\,.
\end{equation} 
In \cref{thermo}, $g_N=2$ is degeneracy of nucleons 
and $E^{\ast }(k)=\sqrt{M_i^{\ast 2}+k^{2}}$. We can relate the quantity $\nu_B$ to the baryon chemical potential $\mu_B$ as \cite{wang}
\begin{align}
\nu_B = \mu_B - g_{\omega}^q\omega -g_{\rho}^q\rho,~~~~~~~~~~~~~~~~~~(q=u,d,s).
\end{align}
The equations of motion for scalar fields $\sigma$, $\zeta$, the dilaton field, $\chi$, scalar iso-vector field, $\delta$, and, the vector fields $\omega$ and $\rho$ are calculated from thermodynamical potential and are written as 
\begin{align}
\label{eq_sigma}
k_{0}\chi^{2}\sigma
-4k_{1}\left( \sigma^{2} \, + \, \zeta^{2} \, + \, \delta^2\right) \sigma \, - \,
2k_{2}\left(\sigma^{3} + 3\sigma\delta^2\right)\, - \, 2k_{3}\chi \sigma \zeta \, - \,
\frac{\xi }{3}\chi^{4}\left(\frac{2\sigma}{\sigma^2-\delta^2}\right)
 \nonumber \\
 \, + \, \frac{\chi^{2}}{\chi _{0}^{2}}m_{\pi }^{2}f_{\pi }
-\left( \frac{\chi }{\chi _{0}}\right)^{2}m_{\omega }\omega ^{2}
\frac{\partial m_{\omega }}{\partial \sigma }\, -\left( \frac{\chi }{\chi _{0}}\right)^{2}m_{\rho }\rho ^{2}
\frac{\partial m_{\rho }}{\partial \sigma }\, + \,
 \sum_{N=p,n} \frac{\partial M_{N}^{\ast }}
{\partial \sigma } \rho_N^s=0,~~~~~
\end{align}
\begin{align}
\label{zeta}
k_{0}\chi^{2}\zeta - 4k_{1}\left(\sigma^{2} \, + \,
\zeta ^{2} + \delta^2\right)
 \zeta \, &- \, 4k_{2}\zeta ^{3} \, - \,
k_{3}\chi \sigma ^{2} \, 
- \, \frac{\xi\chi^{4}}{3\zeta } \, 
\nonumber \\&+ \,
\frac{\chi^{2}}{\chi _{0}^{2}} \left( \sqrt{2}m_{K}^{2}f_{K} \, - \,
\frac{1}{\sqrt{2}}m_{\pi }^{2}f_{\pi } \right)=0,
\end{align}
\begin{align}
\label{scalar1}
 k_0\chi
\left(\sigma^2+\zeta^2+\delta^2\right)
-k_3\left(\sigma^2-\delta^2\right)\zeta+\frac{2\chi}{\chi _{0}^{2}}\left[  m_{\pi }^{2}f_{\pi }\sigma+\left( \sqrt{2}m_{K}^{2}f_{K} \, - \,
\frac{1}{\sqrt{2}}m_{\pi }^{2}f_{\pi } \right)\zeta\right] \nonumber \\
-\frac{\chi}{\chi_0^2} 
\left(m_\omega^2\omega^2+m_\rho^2\rho^2\right)
 +\left(4k_4+1+{\rm ln}\frac{\chi^4}{\chi_0^4} -
\frac{4\xi}
3{\rm ln}\left(\left(\frac{\left(\sigma^2-\delta^2\right)\zeta}{\sigma_0^2\zeta_0}\right)\right)\right)\chi^3 
 =0,
\end{align}
\begin{align}
\label{eq_delta}
k_{0}\chi^{2}\delta
-4k_{1}\left( \sigma^{2} \, + \, \zeta^{2} \, + \, \delta^2\right) \delta \, - \,
2k_{2}\left(\delta^{3} + 3\sigma^2\delta\right)\, + \, 2k_{3}\chi \delta \zeta \, - \,
\frac{\xi }{3}\chi^{4}\left(\frac{2\delta}{\sigma^2-\delta^2}\right)
 \nonumber \\
  + \,
 \sum_{N=p,n}
  g_{\delta N} \rho_N^s=0,~~~~~
\end{align}
\begin{align}
 \left(\frac{\chi^2}{\chi_0^2}\right) 
m_\omega^2\omega+4g_4
\omega^3+12g_4\omega\rho^2-\sum_{N=p,n} g_{\omega N} \rho_N=0, \label{vector1}
\end{align}
and
\begin{align}
 \left(\frac{\chi^2}{\chi_0^2}\right) 
m_\rho^2\rho+4g_4
\rho^3+12g_4\omega^2\rho+\sum_{N=p,n} {g_{\rho N}} \rho_N=0, \label{vector2}
\end{align}
respectively. In \cref{eq_delta,vector1,vector2}, $g_{\delta N}$ is nucleon-delta meson coupling constant and $g_{\omega N}$, $g_{\rho N}$ are nucleon-vector meson coupling constants generated from quark- meson interaction terms in \cref{quarkmesons}. These coupling constants satisfy the SU(3) relationships : $g_{\delta n}=-g_{\delta p}$ and $g_{\rho p}=-g_{\rho n}=\frac{1}{3} g_{\omega p}=\frac{1}{3} g_{\omega p}$.   
In \cref{eq_sigma}, $\rho_i^s$ is the scalar density of nucleons and is  given by
\begin{align}
\rho_N^s =\frac{g_N \,}{2\pi^2} \,
\int_{0}^{{\infty}} dk \frac{k^2 M_N^\ast}{\sqrt{M_N^{\ast 2}+k^2}}\left[n_n(k)
+ \bar n_n(k)+n_p
(k)+ \bar n_p(k)\right]. 
\end{align}
The vector density $\rho_N$ in \cref{vector1,vector2} is given as
\begin{align}
\rho_N =\frac{g_N \,}{2\pi^2} \,
\int_{0}^{{\infty}} dk k^2 \left[n_n(k)
- \bar n_n(k)+n_p
(k)- \bar n_p(k)\right] ,
\end{align}
where, $n_n(k)$ and $n_p(k)$ are the neutron and proton distributions. The anti-neutron and anti-proton distributions  $\bar n_n(k)$ and $\bar n_p(k)$, are respectively defined as
\begin{align}
n_{N}(k)=\lbrace \text{exp}\left[ \left( E^{\ast }(k)-\nu_{B} \right) /k_{B}T \right]+1\rbrace^{-1},~~~~~ 
\label{nucleondis}
\end{align}
and
\begin{align}
\bar n_{N}(k)=\lbrace \text{exp}\left[ \left( E^{\ast }(k)+\nu_{B} \right) /k_{B}T \right]+1\rbrace^{-1},~~~~(N= n,p).
\label{antinucleondis} 
\end{align}
In order to study asymmetric nuclear matter at different values of baryonic density of the medium, we introduce the asymmetry parameter $I$ given as : $I=\frac{\rho_n-\rho_p}{2\rho_B}$, where $\rho_n$ and $\rho_p$ are number densities of neutrons and protons, respectively, and $\rho_B$ is total number density of nuclear medium. 

\subsection{Magnetic Moment of Baryons}
\label{masscalcub} 
In this section we will calculate medium modified magnetic moments of octet baryons in asymmetric nuclear matter using the idea of chiral quark model initiated by Weinberg \cite{sw} and developed by Manohar and Georgi \cite{am}. Although, the magnetic moments can be calculated using hadronic electromagnetic current making use of the wave function of chiral SU(3) quark mean field model as is done in quark meson coupling model calculations in \cite{ryu,ryu3} using the approach in similar lines to \cite{qmcmm}, but using this type of approach one can only calculate medium modification of valence quarks. However, in order to include contributions to magnetic moments of baryons from sea quark effect and that from the orbital angular momentum of sea quarks the approach of chiral quark model is best suited. 

This approach has already been successful for the calculation of medium modified magnetic moments of baryons in symmetric nuclear matter \cite{happy}, where 
the effects of sea quark polarization as well as orbital angular momentum of sea quarks along with the contribution from valence quarks \cite{cheng,cqm} were included. 
As already discussed in \cite{happy} the massless quarks acquire  mass through spontaneous breaking of chiral symmetry in chiral quark model 
through the emission of massless Goldstone boson (GB) identified as ($\pi,K,\eta$) mesons. For calculations within the QCD confinement scale and taking the chiral symmetry breaking into account, one should consider the constituent quarks, GBs and the weakly interacting gluons as the appropriate degrees of freedom \cite{aarti}. The effective Lagrangian in this region is given as
\be
\label{int1} 
{\cal L}_{{\rm interaction}} = \bar{\psi}\left(i{\slashed D} + {\slashed V}\right)\psi +
i g_{A} \bar{\psi} {\slashed A}\gamma^{5}\psi + \cdots \,, 
\ee
where $g_A$ is axial vector coupling constant. In the low energy regime gluonic degrees can be neglected. Hence, the effective interaction Lagrangian in \cref{int1}, by taking GBs and quarks in leading order can be written as
\be
\label{int2} 
{\cal
L}_{{\rm interaction}} = -\frac{g_{A}}{f_{\pi}} \bar{\psi} \partial_{\mu}
\Phi^{'} \gamma^{\mu} \gamma^{5} \psi \,, 
\ee 
where $\Phi^{'}$ represents the octet of GBs given as
\bea 
\resizebox{.5\hsize}{!}{ 
$\Phi^{'} = \left( \ba{ccc} \frac{\pi^o}{\sqrt 2}
+\varpi\frac{\eta}{\sqrt 6} & \pi^+
  & \varepsilon K^+   \\
\pi^- & -\frac{\pi^o}{\sqrt 2} +\varpi \frac{\eta}{\sqrt 6}
 &  \varepsilon K^o  \\
 \varepsilon K^-  &  \varepsilon \bar{K}^o  &  -\varpi \frac{2\eta}{\sqrt 6}
  \ea \right).$ } 
 \nonumber \\
 \eea
 In above, $\varepsilon$ and $\varpi$ are the SU(3) symmetry breaking parameters. The $\text{SU(3)}$ symmetry breaking is introduced by considering
$m_s > m_{u,d}$, as well as by considering
the masses of GBs to be nondegenerate
 $(m_{K,\eta} > m_{\pi})$ {\cite{{chengsu3},{cheng1},{song},{johan}}}. Physically, $\varepsilon^2a$ and $\varpi^2a$ respectively denote the probabilities of transitions $u(d)\rightarrow s+K^-$ and $u(d,s)\rightarrow u(d,s)+\eta$. 
Using the Dirac's equation, i.e., $(i \gamma^{\mu} \partial_{\mu} - m_q)q =0$, the quark-GB interaction Lagrangian in \cref{int2} can be further reduced as \cite{aarti}
\be
\label{int3} 
{\cal
L}_{{\rm interaction}} = i \sum_{q=u,d,s} {g_{8}} \bar{q} 
\Phi^{'}  \gamma^{5} q \,, 
\ee 
where $g_8$ is the coupling constant between octet GBs and quark mass parameter. Supressing all the space-time structure to lowest order the effective Lagrangian 
can be expressed as

\be
{\cal L_{\rm interaction}} = g_8 \bar q \Phi^{'} q\,.
\ee
Further, it is worth noting that the invariance of QCD Lagrangian under axial U(1) symmetry leads to generation of ninth GB realized as $\eta^{'}$. This leads to interaction of quarks with a nonet of GB 
in place of octet of GBs. Thus, interaction Lagrangian can be further written as
\be
\label{int3}
{\cal L_{\rm interaction}} = g_8 \bar q \Phi^{'} q + g_1 \bar q \frac{\eta^{'}}{\sqrt{3}} q =  g_8 \bar q \Phi q,
\ee  
  where $g_1$ is the coupling constant for the singlet and quarks.
Further, in \cref{int3} 
\bea 
q =\left( \ba{c} u \\ d \\ s \ea \right) ,
\eea
and
\bea
\resizebox{.7\hsize}{!}{ 
$\Phi = \left( \ba{ccc} \frac{\pi^o}{\sqrt 2}
+\varpi\frac{\eta}{\sqrt 6}+\tau\frac{\eta^{'}}{\sqrt 3} & \pi^+
  & \varepsilon K^+   \\
\pi^- & -\frac{\pi^o}{\sqrt 2} +\varpi \frac{\eta}{\sqrt 6}
+\tau\frac{\eta^{'}}{\sqrt 3}  &  \varepsilon K^o  \\
 \varepsilon K^-  &  \varepsilon \bar{K}^o  &  -\varpi \frac{2\eta}{\sqrt 6}
 +\tau\frac{\eta^{'}}{\sqrt 3} \ea \right).$} 
 \nonumber \\
 \eea

 The parameter $\tau=g_1/g_8$. Physically $\tau^2a$  denote the probability of transition $u(d,s)\rightarrow u(d,s)+\eta^{'}$. In accordance with NMC (New Muon Collaboration) calculations {\cite{nmc}} we have used 
\begin{equation}
\tau=-0.7-\frac{\varpi}{2} \label{zetadash} .
\end{equation}  
The fluctuation process for describing describing interaction Lagrangian is 
\be
  q_{\pm} \rightarrow {\rm GB}
  + q^{'}_{\mp} \rightarrow  (q \bar q^{'})
  +q_{\mp}^{'}\,.                              \label{basic}
\ee
 Thus the GB further splits into $q \bar q^{'}$ and $q \bar q^{'}  +q_{\mp}^{'}$ constitute the `sea quark'
  \cite{{chengsu3},{cheng1},{song},{johan}}.
The interaction of pion and baryons at one pion loop level can be represented in terms of symmetry breaking parameters $\varepsilon$ and $\varpi$ \cite{oneloop} using the relations 
for octet isospin  multiplets as 
\begin{eqnarray}
&\varpi_N=1,~~~ \varpi_{\Lambda}=\frac{4\varepsilon^2}{3}, ~~~ 
&\varpi_{\Sigma}=4(1-\varepsilon)^2 , ~~~ \varpi_{\Xi}=(1-2\varepsilon)^2. 
\end{eqnarray}
In isospin asymmetric matter the parameter $\varepsilon$ can be defined in terms of effective masses of quarks and baryon (defined in \cref{qmass,baryonmass}) using the relations in Ref. \cite{ling} as
\begin{equation}
\varepsilon=\frac{M_{p}^{*}-M_{n}^{*}}{m_u^*-m_d^*}, \label{alphadash1}
\end{equation}
whereas in isospin symmetric limit ($m_u^*=m_d^*$) the above definition of $\varepsilon$ turns to be  
\begin{equation}
\varepsilon=\frac{M_{\Sigma}^{*}-M_{\Xi}^{*}}{\left(\frac{m_{u}^*+m_{d}^*-2m_{s}^*}{2}\right)}. \label{alphadash}
\end{equation} 

The total effective magnetic moment of baryons 
can be written as
 \begin{align}
 \mu\left( B\right)_{\text{total}}= \mu\left( B\right)_{\text{val}}+\mu\left( B\right)_{\text{sea}}+\mu\left( B\right)_{\text{orbital}}, \label{magtotal}
 \end{align}
 where $\mu\left( B\right)_{\text{val}}$, $\mu\left( B\right)_{\text{sea}}$ and $\mu\left( B\right)_{\text{orbital}}$ represent the contribution from valence quarks, sea quarks and orbital angular momentum of sea quarks, respectively. For details the reader may refer to \cite{happy,har,cheng1,song}. For medium modification of these effects, the calculations are explicitly done in \cite{happy}. 

The values of effective magnetic moment of constituent quark ($\mu_q$) can be calculated following the naive quark model formula given as 
$\mu_{\rm q}=\frac{e_{\rm q}}{2m_{\rm q}}$, where $m_{\rm q}$ and $e_{\rm q}$ are mass and electric charge of quark, respectively. This formula lacks consistency for calculation of magnetic moments of relativistically confined quarks \cite{har2} and the non-relativistic quark momenta are required to be very small ($p^2_q<<(350 {\rm MeV})^2$) for quark masses in the range of $313$ MeV and more. Therefore, in order to include quark confinement effect on magnetic moments \cite{har2,gupta} along with relativistic correction to quark magnetic moments (introduced in quarks by using medium modified quark masses obtained in chiral SU(3) quark mean field model, which considers quarks as Dirac particles), the mass term in the formula for quark magnetic moment is replaced by the expectation value of effective quark mass $\bar{m_{\rm q}}$, which can be further expressed in terms of effective baryon mass $M_{\rm B}^*$ (B = $p,n,\Sigma^{+},\Sigma^{0},\Sigma^{-},\Xi^{0},\Xi^{-},\Lambda$) following the formula 
\begin{eqnarray}
2\bar{m_{\rm q}}=M_{\rm B}^*+m_q \nonumber,
\end{eqnarray} 
$m_q(\approx 0)$ is the current quark mass. 
To include quark confinement effect, $M_{\rm B}^*$ is replaced by $M_B^*+\Delta M$. $\Delta M$ being the difference between experimental vacuum mass of baryon ($M_{\rm vac}$) and the effective mass of baryon $M_B^*$,($\Delta M=M_{\rm vac}-M_{B}^{*}$) \cite{har2}. The mass difference $\Delta M$ represents the indivisible and irreducible part to confining energy \cite{nobo}. 

Following the above formalism, the equations 
to calculate the effective magnetic moments `${\mu_q}$' of constituent quarks are now given as
\begin{equation}
 \mu_d =-\left(1-\frac{\Delta M}{M_B^*}\right),~~  \mu_s=-\frac{m_u}{m_s}\left(1-\frac{\Delta M}{M_B^*}\right),~~  \mu_u=-2\mu_d .\label{magandmass}         
\end{equation}     
These 
are referred to as the mass adjusted magnetic moments of constituent quarks \cite{aarti}. $M_B^*$ is obtained in equation (\ref{baryonmass}). The choice of these definitions of quark magnetic moments make it possible to study their medium modification directly from the in-medium masses not only for the constituent quarks masses but also for the in-medium baryon masses. 
The orbital angular momentum contribution is calculated using the parameters $\varepsilon,\varpi$ and $\tau$ along with masses of GBs using the details in Ref. \cite{happy}. The GBs contributions are dominated by pion contribution as compared to contributions from other GBs. 
 The medium modification of parameter $\varepsilon$ given by \cref{alphadash1,alphadash} lead to medium modification of sea quark polarizations and orbital angular moments.
\section{Numerical results} \label{results}
\begin{figure}
\begin{center}
\includegraphics[scale=.58]{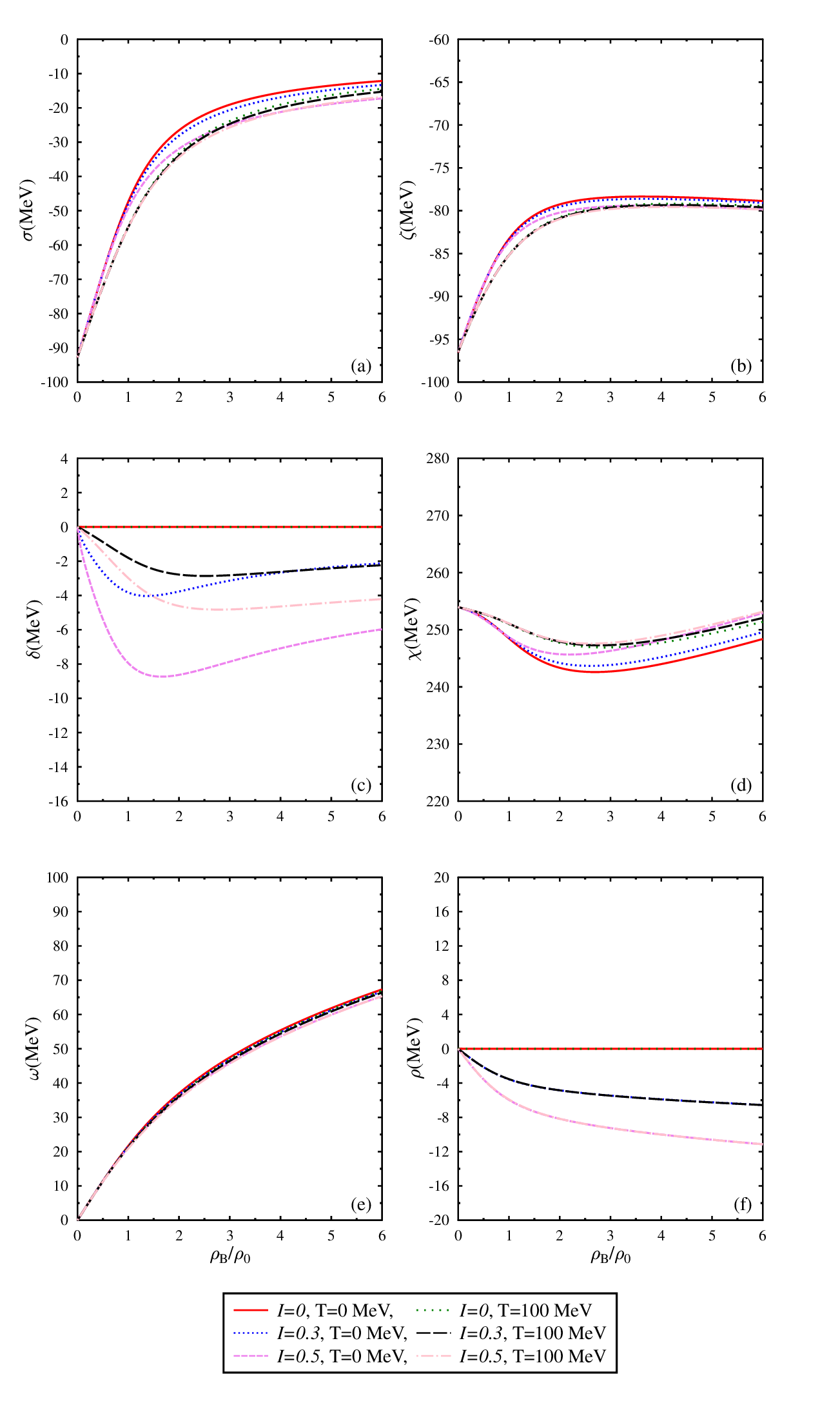} 
\caption{{$\sigma$, $\zeta$, $\omega$, $\chi$, $\delta$ and $\rho$ fields (at T = 0 and 100 MeV and asymmetry parameters $I=0,0.3$ and $0.5$) versus baryonic density (in units of nuclear saturation density $\rho_0$).} } \label{fields}
\end{center}   
\end{figure}

In this section we present the results of our investigation on magnetic moment of baryons at finite density and temperature of the asymmetric nuclear medium. Various parameters used in the present work are listed in \cref{cc}. From \cref{magandmass} it is clear that the value of magnetic moment of constituent quarks depends on the effective masses of the quarks and baryons, which in-turn depends on the scalar fields $\sigma$ and $\zeta$ and the scalar isovector field $\delta$ through \cref{qmass}. In order to study the dependence of the density of medium at different values of temperature `T' and 
isospin asymmetry parameter `$I$' on the scalar fields $\sigma$ and $\zeta$ and the scalar isovector field $\delta$ along with other meson fields, in figs. \ref{fields}(a)-(f), we plot the fields $\sigma,\zeta,\delta,\chi,\omega$ and $\rho$ as a function of nuclear matter density $\rho_B$ (in the units of nuclear saturation density $\rho_0$), at temperatures T = 0 and 100 MeV and asymmetry parameters $I=0, 0.3$ and $0.5$.
\begin{table}
\begin{tabular}{|c|c|c|c|c|}
\hline 
$g_{\delta p}$  & $g_{\omega p}$  & $g_{\rho p}$  & $m_{\pi}$ (MeV) & $m_K$ (MeV) \\ 
\hline 
2.5 & 11.59 & 3.86 & 139 & 494  \\ 
\hline
\hline 
\hline 
$k_0$ & $k_1$ & $k_2$ & $k_3$ & $k_4$  \\ 
\hline 
4.94 & 2.12 & -10.16 & -5.38 & -0.06  \\ 
\hline
\hline
\hline 
$\sigma_0$ (MeV) & $\zeta_0$ (MeV) & $\chi_0$ (MeV) & $\xi$ & $\rho_0$ ($\text{fm}^{-3}$)  \\ 
\hline 
-92.8 & -96.5 & 254.6 & 6/33 & 0.16  \\ 
\hline
\hline
\hline 
$g_{\sigma}^u=g_{\sigma}^d$ & $g_{\sigma}^s$ & $g_{\zeta}^u=g_{\zeta}^d$ & $g_{\zeta}^s$ & $g_4$\\ 
\hline 
3.37 & 0 & 0 & 4.77 & 37.4 \\ 
\hline
\end{tabular}
\caption{Values of various parameters used in the present work \cite{wang}.} \label{cc}
\end{table} 
\begin{figure}
\begin{center}
\includegraphics[scale=.58]{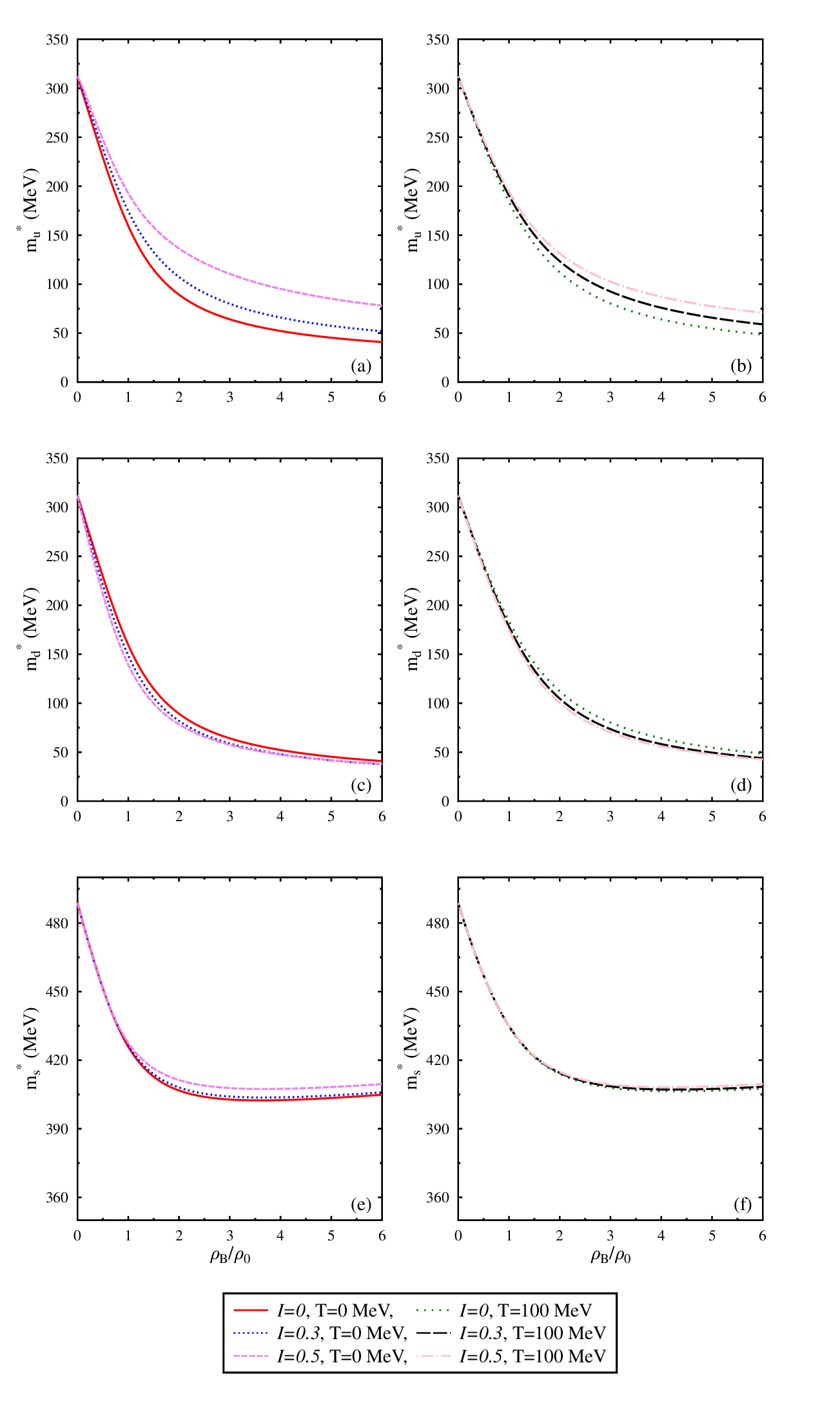} 
\caption{{Effective masses of quarks (at T = 0 MeV (in left panel) and T = 100 MeV (in right panel) and asymmetry parameters $I=0, 0.3$ and $0.5$) versus baryonic density (in units of nuclear saturation density $\rho_0$).} } \label{qmasst}
\end{center} 
\end{figure}
\begin{figure}
\begin{center}
\includegraphics[scale=.7]{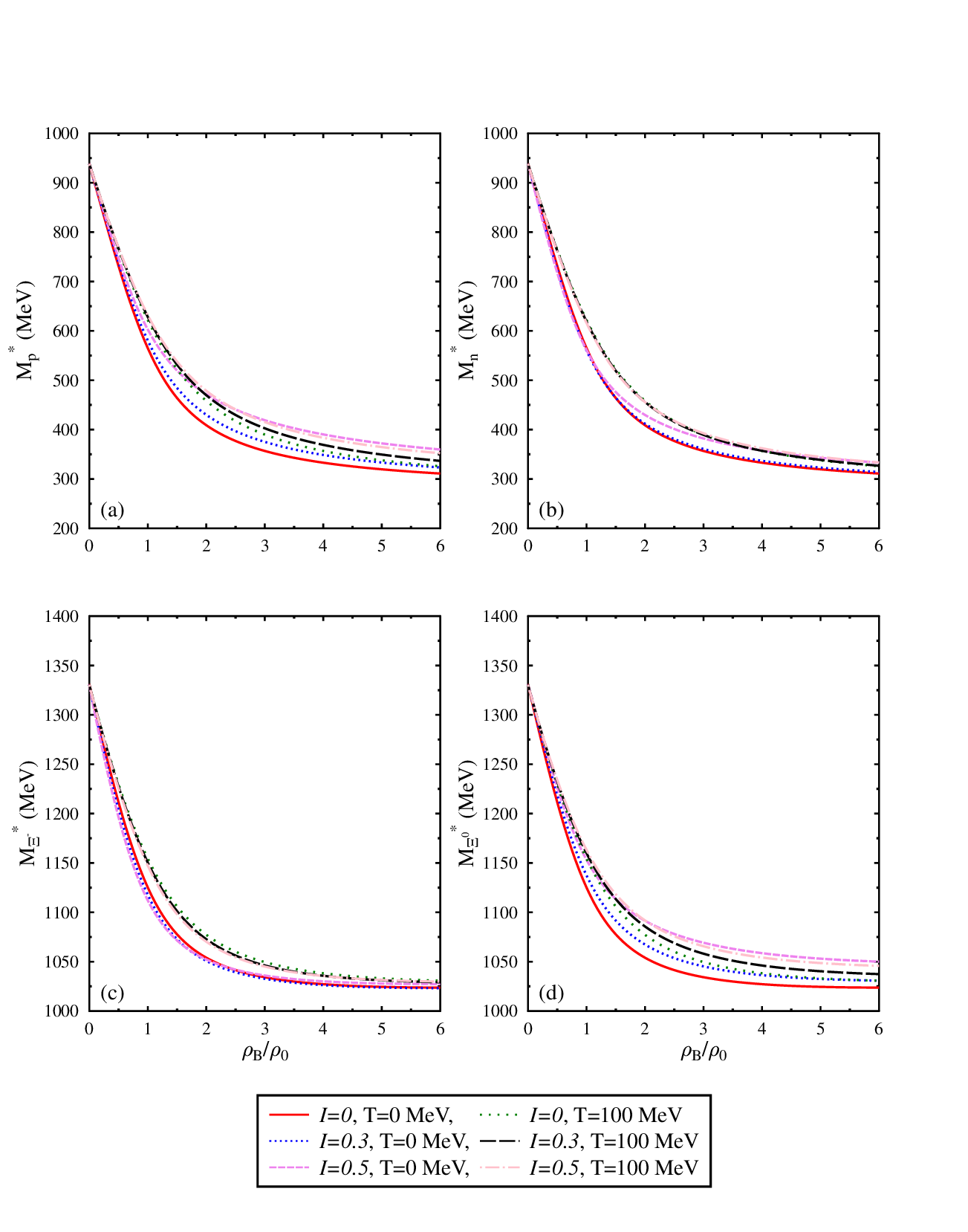} 
\caption{{Effective masses of $p,n,\Xi^{-}$ and $\Xi^{0}$ baryons (at T=0  and 100 MeV and asymmetry parameters $I=0,0.3$ and $0.5$) versus baryonic density (in units of nuclear saturation density $\rho_0$).} } \label{bmasst1}
\end{center} 
\end{figure}
\begin{figure}
\begin{center}
\includegraphics[scale=.7]{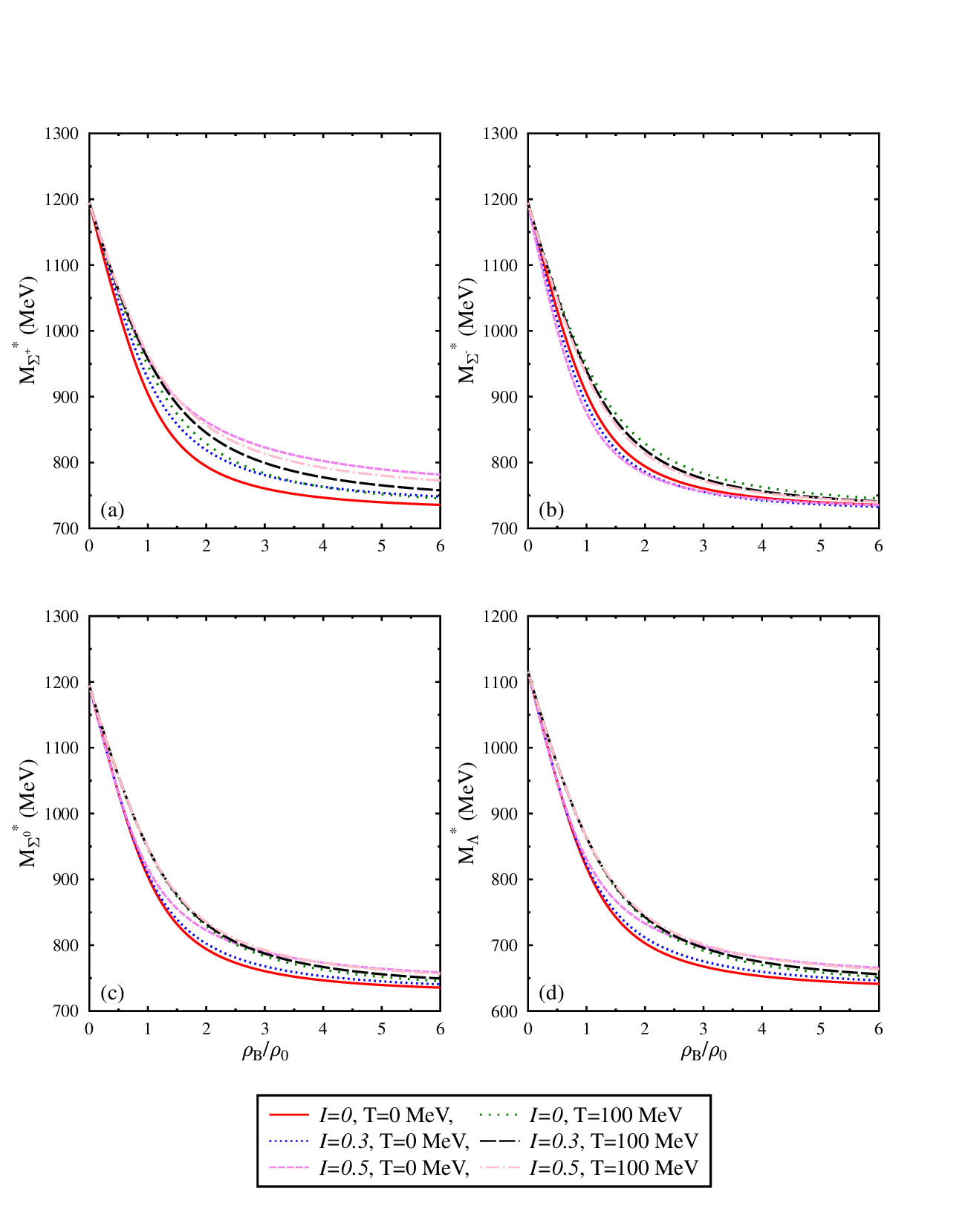} 
\caption{{Effective masses of $\Sigma^{+},\Sigma^{-},\Sigma^{0}$ and $\Lambda$ baryons (at T=0  and 100 MeV and asymmetry parameters $I=0,0.3$ and $0.5$) versus baryonic density (in units of nuclear saturation density $\rho_0$).} } \label{bmasst2}
\end{center} 
\end{figure}
\begin{figure}
\begin{center}
\includegraphics[scale=.75]{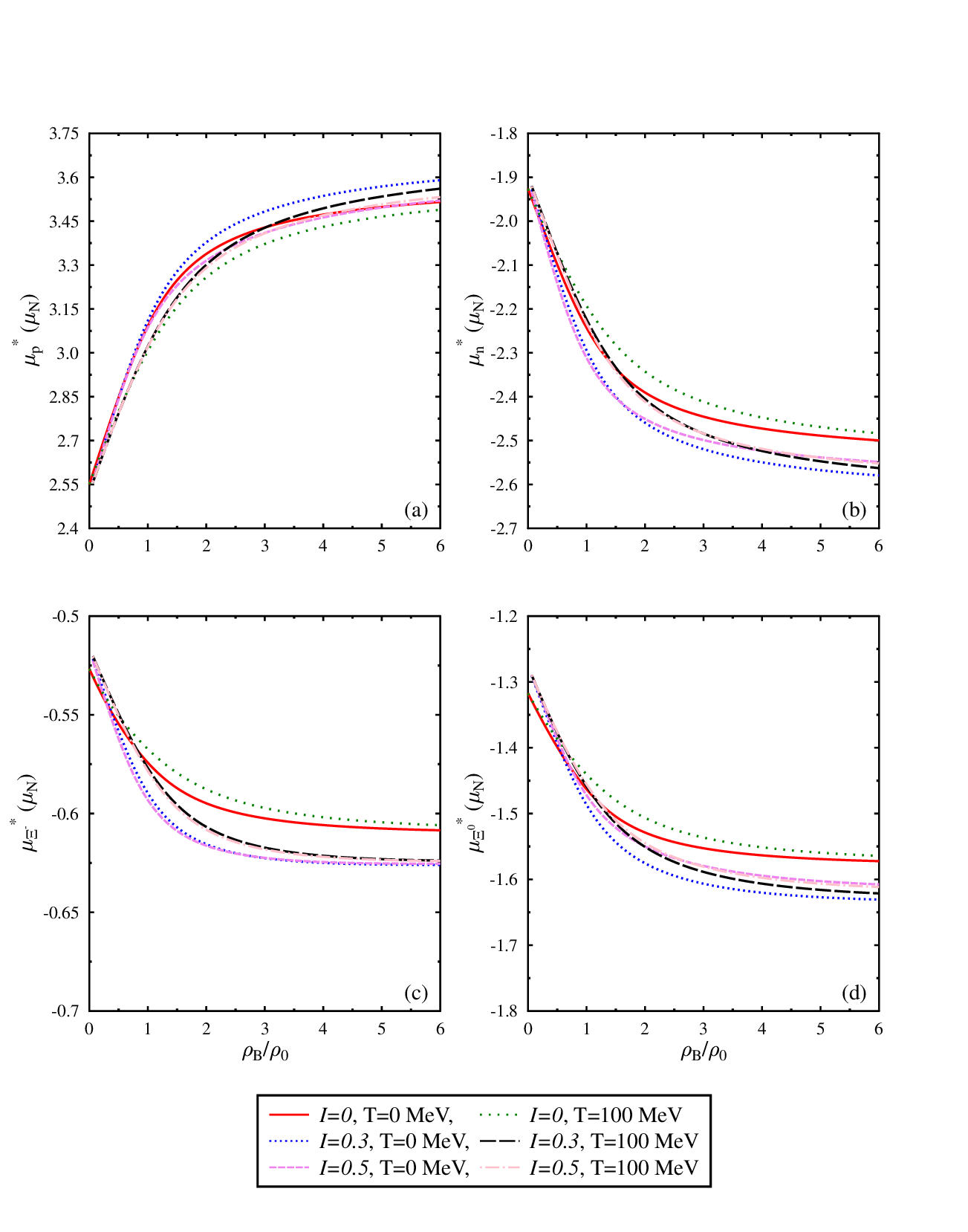} 
\caption{{Magnetic moment of $p,n,\Xi^{-}$ and $\Xi^{0}$ baryons (at T = 0 and 100 MeV and asymmetry parameters $I=0,0.3$ and $0.5$) versus baryonic density (in units of nuclear saturation density $\rho_0$).} } \label{mmt1}
\end{center} 
\end{figure} 
\begin{figure}
\begin{center}
\includegraphics[scale=.75]{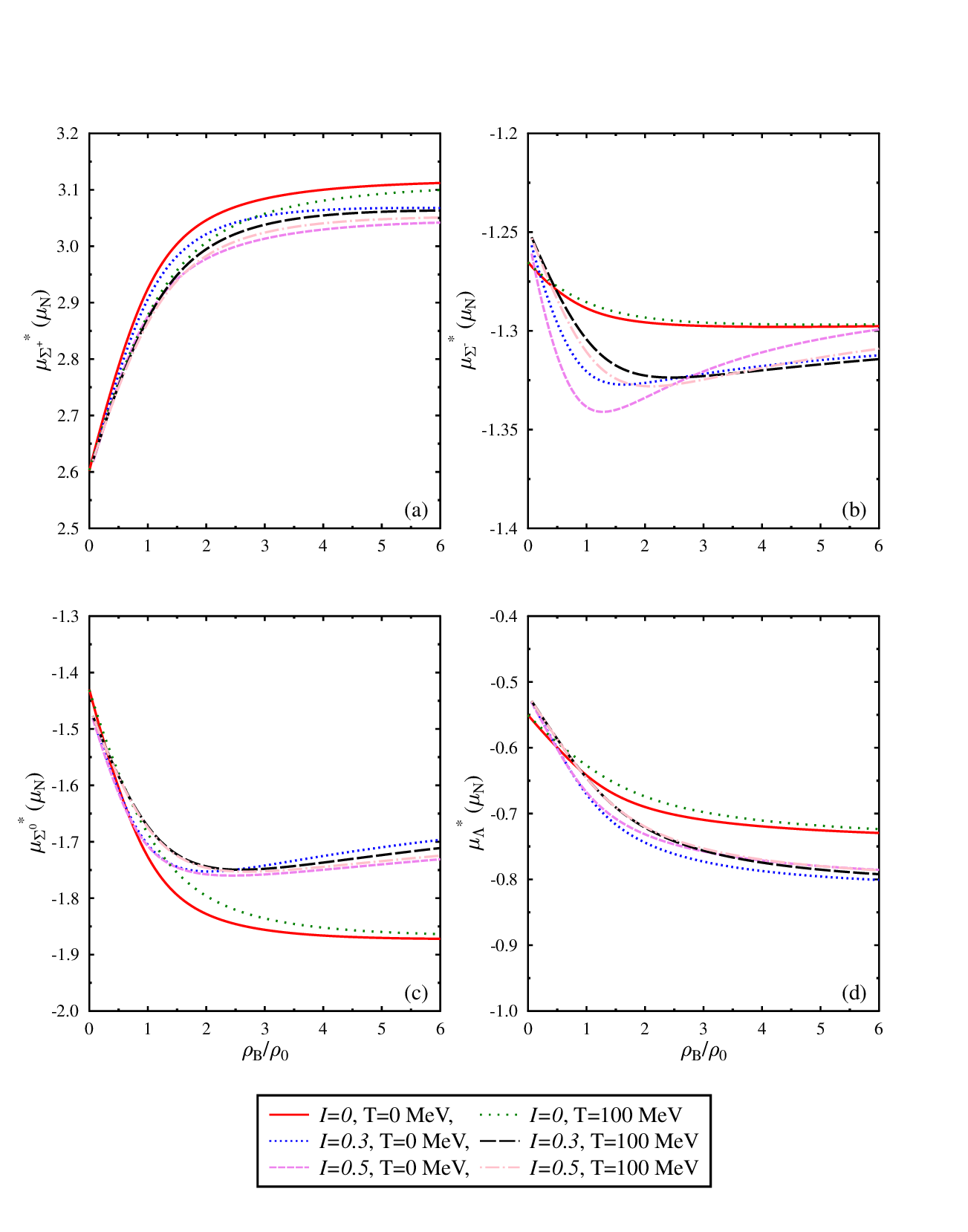} 
\caption{{Magnetic moment of $\Sigma^{+},\Sigma^{-},\Sigma^{0}$ and $\Lambda$ baryons (at T = 0 and 100 MeV and asymmetry parameters $I=0,0.3$ and $0.5$) versus baryonic density (in units of nuclear saturation density $\rho_0$).} } \label{mmt2}
\end{center} 
\end{figure}
\begin{figure}
\begin{center}
\includegraphics[scale=.75]{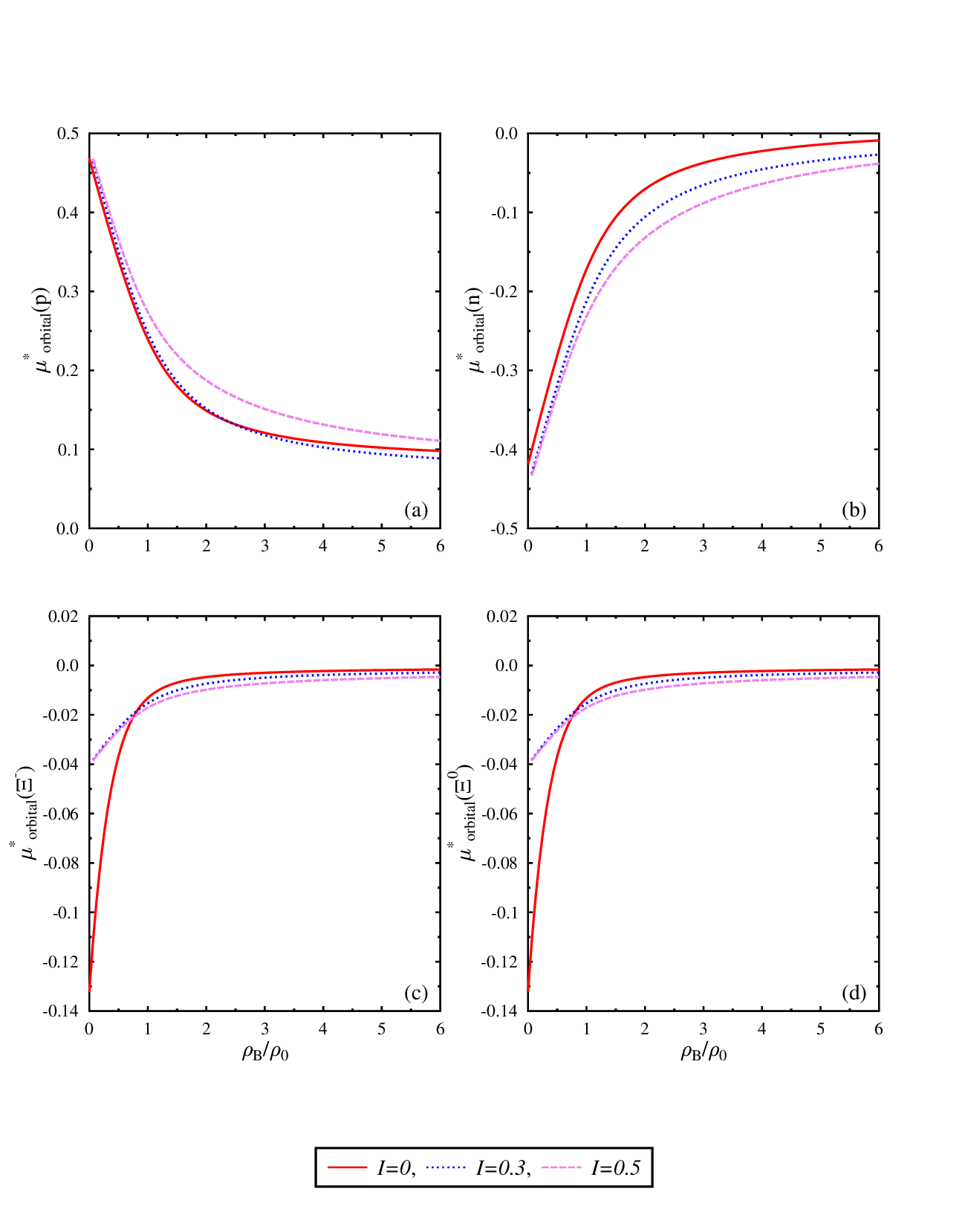} 
\caption{{Contribution from orbital angular momentum of sea quarks to the total magnetic moment of $p,n,\Xi^{-}$ and $\Xi^{0}$ baryons (at T = 0 MeV and asymmetry parameters $I=0,0.3$ and $0.5$) in the units of $\mu_N$ versus baryonic density (in units of nuclear saturation density $\rho_0$).} } \label{ommt1}
\end{center} 
\end{figure} 
\begin{figure}
\begin{center}
\includegraphics[scale=.75]{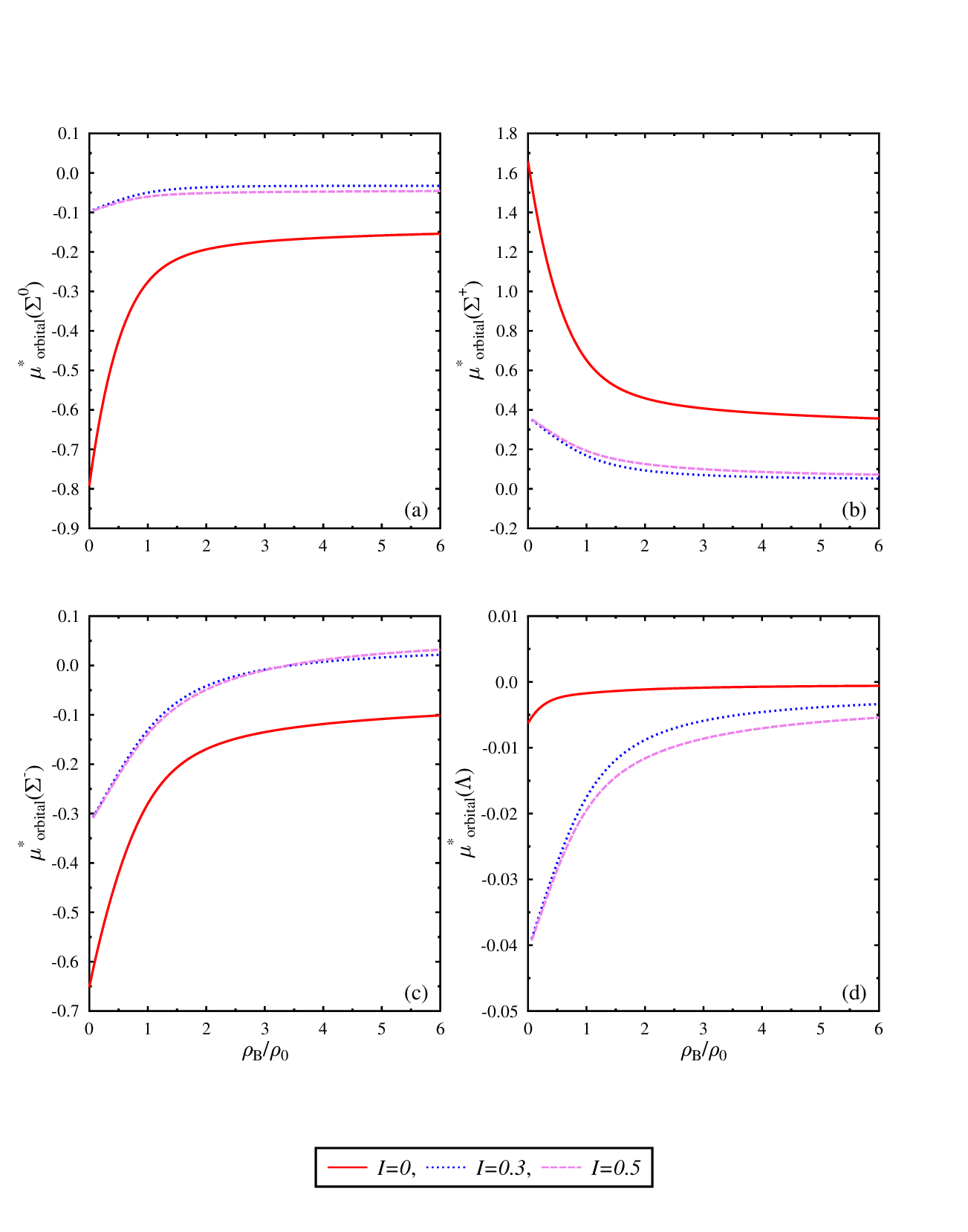} 
\caption{{Contribution from orbital angular momentum of sea quarks to the total magnetic moment of $\Sigma^{+},\Sigma^{-},\Sigma^{0}$ and $\Lambda$ baryons (at T = 0 MeV and asymmetry parameters $I=0,0.3$ and $0.5$) in the units of $\mu_N$ versus baryonic density (in units of nuclear saturation density $\rho_0$).} } \label{ommt2}
\end{center} 
\end{figure}

We observe that for a given value of asymmetry parameter and temperature, the magnitude of scalar $\sigma$ field decreases sharply with the rise of nuclear matter density from $\rho_B=0$ to $2\rho_0$ as compared to magnitude of $\zeta$ field. This is due to strong correlation between the nucleons and the $\sigma$ field, whereas, 
the absence of coupling between $\zeta$ field and non-strange quark content of the medium leads to slow decrease in its magnitude as a function of density. For densities more than $2\rho_0$, the decrease in magnitude of scalar fields as a function of $\rho_B$ is slow. For example, at T = 0 MeV and $I=0, 0.3$ and $0.5$, the magnitude of $\sigma$ field decreases by $73\%$, $66\%$ and $62\%$ from its vacuum value, respectively, as $\rho_B$ increases from $\rho_B=0$ to $2\rho_0$, whereas, there is a decrease of only about $17\%$ at $I=0$ and $I=0.3$ and about $16\%$ at $I=0.5$, in magnitude of $\zeta$ field. This also shows that for given temperature and density of medium, magnitude of $\sigma$ field increases with increase of isospin asymmetry of medium. However, $\zeta$ field is negligibly affected by isospin asymmetry of medium. 
For given isospin asymmetry and density of the medium, the magnitude of scalar fields $\sigma$ and $\zeta$ are observed to increase with increase of temperature. For example, for $\rho_B=\rho_0$ and $I=0.3$, the magnitude of $\sigma$ field are 48.04 MeV and 54.8 MeV, at T = 0 and 100 MeV, respectively.
Further, for given density of the medium, at finite temperature, $\sigma$ field increases less with increase of asymmetry parameter $I$ as compared to T = 0 MeV case. 
This is because with the rise of temperature the thermal distribution function given in \cref{nucleondis,antinucleondis} start decreasing, which lead to soft equation of state at finite temperatures than at zero temperature. Due to this the coupling strength between $\sigma$ field and nucleons decrease with the rise of temperature. 
The scalar isovector $\delta$ field plotted in \cref{fields}(c) contributes in isospin asymmetric medium only. 
In symmetric nuclear matter ($I=0$), $\delta$ field remains $0$ MeV even with the increase of density and temperature of the medium. For given temperature and isospin asymmetry of the medium, 
the magnitude of $\delta$ field starts increasing with the rise of density from $\rho_B=0$ to $2\rho_0$ and then it decreases to certain value, upto $6\rho_0$. For example, at T = 0 MeV and $I=0.3$, magnitude of $\delta$ field increases upto $4.054$ MeV at $\rho_B=1.5\rho_0$ and then it decreases to $2.38$ MeV at $\rho_B=6\rho_0$. Similarly, at $I=0.5$, $\delta$ field acquires its highest magnitude $8.82$ MeV at $\rho_B=1.8\rho_0$ and its magnitude decreases to $6.6$ MeV at $\rho_B=6\rho_0$.
 
 At T = 100 MeV, the $\delta$ field shows a similar variation as a function of density as that at T = 0 MeV. However, magnitude of $\delta$ field has its maximum value at higher value of density of medium as compared to the case of T = 0 MeV. For example, at $I=0.3$, $\delta$ field acquires its maximum magnitude $2.946$ MeV at $\rho_B=3\rho_0$ and then its magnitude decreases to $2.466$ MeV at $6\rho_0$. Further, at $I=0.5$, magnitude of $\delta$ field increases to $5.013$ MeV at $\rho_B=3\rho_0$ and then it decreases to $4.608$ MeV at $6\rho_0$. Thus, $\delta$ field varies more as a function of density at T = 0 MeV than that at T = 100 MeV for given finite value of asymmetry parameter. 
  
  The dilaton field $\chi$ plotted in \cref{fields}(d) is introduced in the model to mimic trace anomly in QCD at tree level \cite{wang}. One can see that at zero as well as finite temperature, the magnitude of $\chi$ field first decreases 
upto certain value of density of medium and then it increases with rise of density. We see that $\chi$ field has least variation as a function of density as compared to other meson fields. Due to this reason in most of mean field studies frozen glueball limit is assumed, i.e., $\chi$ field is taken to be constant. 

 
We observe that isoscalar vector field $\omega$ plotted in \cref{fields}(e) increases almost linearly with the rise of density at zero as well as finite temperature. The variation of $\omega$ field as a function of density is negligibly effected by the increase of isospin asymmetry  upto $3\rho_0$, at zero as well as finite temperature. 
In \cref{fields}(f), we have shown the variation of vector isovector $\rho$ field. 
This field plays important role in the study of asymmetric nuclear matter because it provides repulsive potential between the nucleons in the isovector channel, opposite to that provided by the $\delta$ meson field. 
For given density of the medium, the $\rho$ field does not vary significantly as a function of temperature. However, for given temperature and finite isospin asymmetry of the medium, 
the $\rho$ field decreases as a function of density, at zero as well as finite temperature. 
     
Using the above calculated values of scalar fields $\sigma$, $\zeta$ and scalar isovector $\delta$ field, the in-medium quark masses, $m_q^*$, can be evaluated using \cref{qmass}. In \cref{qmasst}, we plot the medium modified quark masses $m_q^*$ ($q=u,d,s$) as a function of nuclear matter density, at temperature T = $0$ and $100$ MeV and asymmetry parameters  $I=0,0.3$ and $0.5$. 
In case of symmetric as well as asymmetric nuclear matter, 
there is steep decrease in $m_q^*$ at lower values of density upto $2\rho_0$. However, at given value of temperature and asymmetry parameter, the decrease in $m_q^*$ is quite less at more higher densities. The probable cause behind this behavior can be the chiral symmetry restoration at higher densities, which has been reported in literature, by using chiral hadronic model, in the quark degrees of freedom \cite{rau}. 

For given density and temperature of the nuclear medium, the finite isospin asymmetry causes the splitting of $m_u^*$ and $m_d^*$. For example, at $\rho_B=\rho_0$ and T = 0 MeV, as $I$ increases from 0 to 0.3, $m_u^*$ increases by 15.1 MeV and $m_d^*$ decreases by 10.7 MeV as observed in \cref{qmasst}(a) and \cref{qmasst}(c). 
Further, comparing the effective masses at T = 0 and 100 MeV, we find that at finite temperature the effect of isospin asymmetry on effective quark masses is reduced. 
The effective mass of strange quark 
$m_s^*$, decreases less rapidly as a function of density, as compared to $m_u^*$ and $m_d^*$, at given value of temperature and isospin asymmetry parameter. Further, the effect of asymmetry parameter on the variation of $m_s^*$ as a function of density is very small at zero as well as finite temperature. For example, in symmetric matter, at temperature T = 0 and 100 MeV, as the  density of medium increases from $\rho_B=0$ to $\rho_0$, $m_s^*$ decreases from its vacuum value by about $14\%$ and $11\%$, respectively. The reason for this behavior of $m_s^*$ at finite baryonic density is its dependence on scalar $\zeta$ field, and, the absence of coupling between $s$-quark and scalar isovector $\delta$ field. The effect of temperature on effective quark masses has been studied in detail in Ref. \cite{happy}. In the present work, we emphasize on effect of isospin asymmetry of medium.  
We see that the variation of $m_s^*$ as a function of density remains almost un-effected upto $\rho_0$. 
We also observe that in symmetric as well as asymmetric medium, $m_s^*$ decreases with rise of density upto $4\rho_0$, at zero as well as at finite temperature. However, on further increase of density  above $4\rho_0$, at the same given value of temperature and asymmetry parameter, $m_s^*$ starts increasing. 
This variation can be understood from the dependence of $m_s^*$ on value of $\zeta$ field whose magnitude is found to increase at densities more than 4$\rho_0$ as depicted by \cref{fields}(b). Probable cause of this behavior can be  deconfinement phase transition at higher density of the medium \cite{abu}.


Now we discuss the medium modification of octet baryon masses 
calculated using \cref{baryonmass}. 
In figs. \ref{bmasst1} and \ref{bmasst2}, we plot the medium modified octet baryon masses, $M_j^*$ ($j=n, p, \Xi^{-}, \Xi^{0}, \Sigma^{+}, \Sigma^{-}, \Sigma^{0}, \Lambda$), as a function of density at temperatures, T = $0$ and $100$ MeV and asymmetry parameters $I=0,0.3$ and $0.5$. 

In case of nucleons having non-strange quark content only plotted in \cref{bmasst1}(a) and \cref{bmasst1}(b), a steep decrease in the effective masses of nucleons with the rise of density at given temperature and isospin asymmetry parameter is observed. This behavior has been reported also in the Non-linear Walecka model \cite{nldel} with $\delta$ meson and density dependent relativistic mean field model at hadron level in \cite{dd} which is in contrast to a very small decrease reported in a quark meson coupling model calculation \cite{qmcasy}.
For given density and temperature of the nuclear medium, the finite isospin asymmetry of the medium causes the mass splitting in baryon isospin multiplets. In case of nucleon doublet, the mass of proton is observed to increase more as a function of isospin asymmetry as compared to mass of neutron (comparing \cref{bmasst1}(a) and \cref{bmasst1}(b)). For example, in symmetric nuclear matter ($I=0$), the mass of proton and neutron at $\rho_B=3\rho_0$ and T = 0 MeV is observed to be $356.7$ MeV, whereas, at same temperature and density of the medium, with the increase of isospin asymmetry to $I=0.5$ 
the effective masses of proton and neutron are found to be $418$ and $382$ MeV, respectively. The similar behavior has been reported in \cite{miyazaki} by using relativistic mean field model at hadronic level. 
This behavior of effective nucleon masses can be understood in terms of behavior of their constituent quark masses. 
As discussed earlier, for given density and temperature of the medium, $m_u^*$ increases with the increase of isospin asymmetry parameter, whereas, $m_d^*$ decreases. 
In the case of proton, 
the effect of increasing $m_u^*$ with the increase of isospin asymmetry parameter is more dominating and in the case of neutron 
the effect of the increase of effective mass of $u$ quark is less compensated by decrease in mass of two $d$ quarks. Hence, effective mass of neutron increases less with the increase of asymmetry parameter. 
In the present work, the observed mass splitting between $m_u^*$ and $m_d^*$ arises due to introduction of scalar isovector $\delta$ field in \cref{qmass}. Similar effect of inclusion of $\delta$ meson field on the mass splitting between proton and neutron has  been observed in Walecka type models \cite{liu,kubis} and density dependent relativistic hadron model \cite{dd}. The observed mass splitting is however in opposite sense to that  observed in the models solved at Hartee Fock level with $\rho$ field but without the contribution from $\delta$ meson field \cite{qmcasy22,qmcasy23,dd,mass41}. This shows that inclusion of $\delta$ coupling leads to large attractive potential \cite{effmass} than the repulsive potential provided by vector isovector $\rho$ field. 

On comparing values of $M_p^*$ at T = 0 MeV and at 100 MeV, we find that there is more increase of $M_p^*$ with the increase of asymmetry parameter at zero temperature as compared to that at T = 100 MeV. 
In the case of neutron for given temperature and density of medium, 
with the increase of asymmetry parameter upto $I=0.3$, the effective mass increases 
negligibly with the increase of asymmetry of the medium, specially upto $\rho_0$. On further increase in asymmetry parameter to $I=0.5$, 
the effective mass of neutron increases more for the densities higher than $\rho_0$. This shows that in pure neutron matter, isospin asymmetry of the medium starts effecting $M_n^*$ above the saturation density. The cause behind this behavior can be the more compressible nature of pure neutron matter \cite{compress} above saturation density. 
Further, at T = 100 MeV in symmetric medium, there is less decrease in the effective mass of neutron as a function of density as compared to that at T = 0 MeV. 

As compared to non strange baryons, the in-medium masses of strange baryons 
decrease less rapidly as a function of density of medium, at given temperature in symmetric medium as well as in asymmetric nuclear medium. Further, strange baryons with more number of $u$ quarks shows increase in effective mass with increase of isospin asymmetry and baryons with large $d$ quark content show decrease in effective mass with isospin asymmetry. 
In figs. \ref{bmasst2}(a) and (b), we have plotted effective masses of
$\Sigma^+$ and $\Sigma^-$ baryons. 
In case of $\Sigma^-$ 
one finds that, for given density and temperature of medium, with the rise of asymmetry of medium, $M_{\Sigma^-}^*$ decreases. 
For example, at T = 0 MeV and $\rho_B=\rho_0$, $M_{\Sigma^-}^*$ has values 904.6, 888.9 and 875.6 MeV at $I=0,0.3$ and $0.5$, respectively. However, for T = 100 MeV case, the values of $M_{\Sigma^-}^*$ at $I=0,0.3$ and $0.5$ are 947.5, 939  and 933 MeV, respectively. 
On the other hand, $M_{\Sigma^+}^*$ increases 
with the rise of isospin asymmetry parameter, at given temperature and density of medium. Further, one finds that the values of $M_{\Sigma^-}^*$ and $M_{\Sigma^+}^*$ increase 
with the rise of temperature of medium, in symmetric as well as isospin asymmetric nuclear medium. 

In case of $\Sigma^0$ and $\Lambda$ (plotted in figs. \ref{bmasst2}(c) and (d)) 
having different vacuum masses but same quark content, i.e., $uds$, it is observed that their effective masses show similar variation with the rise of density, at given temperature and asymmetry parameter. This is because in the present work the vacuum mass difference between $\Sigma^0$ and $\Lambda$ occurs due to the difference of spin energy $E_{\rm spin}$ for the two baryons which is taken to be constant. In case of $\Sigma^0$ as well as $\Lambda$ baryon, at T = 0 as well 100 MeV, effective mass decreases with the rise of density in symmetric as well as asymmetric matter. Further, at higher value of isospin asymmetry of the medium, the baryons show less decrease in their effective masses as a function of baryonic density. 

In  figs. \ref{bmasst1}(c) and (d), we have plotted effective masses of $\Xi^-$ and $\Xi^0$ baryons. Due to presence of two $s$ quarks and one non strange quark ($u$ or $d$) in $\Xi$ baryons, 
the decrease in the effective masses as a function of density is least as compared to other baryons. 
 Further, with the rise of asymmetry parameter, at given density and temperature of medium, the effective mass of $\Xi^-$ decreases, whereas, effective mass of $\Xi^0$ increases. This is because of dominance of decreasing $m_d^*$ in the case of $\Xi^-$ and increasing $m_u^*$ in the case of $\Xi^0$ with the rise of isospin asymmetry of the medium. 
 

Now we will discuss the medium modification of the magnetic moments of octet baryons. In \cref{mmt1,mmt2}, we plot effective magnetic moments of octet baryons as a function of density at temperatures T = 0 and $100$ MeV and isospin asymmetry parameters $I=0,0.3$ and $0.5$. To understand the explicit dependence of effective magnetic moments on density, temperature of the medium and asymmetry parameter, we have presented the magnetic moments of octet baryons in \cref{mmval} and \cref{mmvaldifft}. 
On comparing the values for $\rho_B=0$ at T = 0 MeV and T = 100 MeV respectively, in \cref{mmval} and \cref{mmvaldifft}, we find that the magnetic moments of baryons are almost same at T = 0 MeV and T = 100 MeV. However, at $\rho_B=\rho_0$, there is a noticeable change in the values of magnetic moments of baryons, with the increase of temperature as well as with the increase of asymmetry parameter.   
 
 \begin{table}
 \begin{tabular}{|c|c|c|c|c|c|c|c|c|}  
 \hline 
 &  \multicolumn{2}{c|}{$\rho_B=0$} & \multicolumn{3}{c|}{$\rho_B=\rho_0$} & \multicolumn{3}{c|}{$\rho_B=4\rho_0$}    \\ 
 \hline
 & Exp. value \cite{experiment} & ${\rm Present Work}$  & $I=0$ & $I=0.3$ & $I=0.5$ & $I=0$ & $I=0.3$ & $I=0.5$  \\ 
 \hline 
 $\mu_p^*(\mu_N)$ & 2.79 & $2.56$  & $3.09$ & $3.11$ & $3.09$ & $3.47$ & $3.54$ & $3.46$   \\ 
 \hline 
 $\mu_n^*(\mu_N)$ & -1.91 & $-1.93$  & $-2.24$ & $-2.29$ & $-2.31$ & $-2.47$ & $-2.55$ & $-2.52$   \\ 
 \hline 
 $\mu_{\Sigma^+}^*(\mu_N)$ & 2.46 & $2.60$ & $2.92$ & $2.91$ & $2.88$ & $3.10$ & $3.06$ & $3.03$    \\ 
 \hline 
 $\mu_{\Sigma^-}^*(\mu_N)$ & -1.16 & $-1.26$ & $-1.29$ & $-1.32$ & $-1.34$ & $-1.30$ & $-1.32$ & $-1.31$   \\ 
 \hline 
 $\mu_{\Sigma^0}^*(\mu_N)$ & -1.61 & $-1.48$ & $-1.73$ & $-1.70$ & $-1.71$ & $-1.87$ & $-1.72$ & $-1.75$   \\ 
 \hline 
 $\mu_{\Xi^0}^*(\mu_N)$ & -1.25 & $-1.32$ & $-1.46$ & $-1.49$ & $-1.47$ & $-1.56$ & $-1.62$ & $-1.59$  \\ 
 \hline 
 $\mu_{\Xi^-}^*(\mu_N)$ & -0.65 & $-0.57$ & $-0.57$ & $-0.58$ & $-0.59$ & $-0.60$ & $-0.62$ & $-0.62$   \\ 
 \hline 
 $\mu_{\Lambda}^*(\mu_N)$ & -0.61 & $-0.55$ & $-0.64$ & $-0.67$ & $-0.67$ & $-0.72$ & $-0.79$ & $-0.77$   \\ 
 \hline 
 \end{tabular} 
\caption{Effective magnetic moments of octet baryons at T = 0 MeV and $\rho_B=0, \rho_0, 4\rho_0$, corresponding to $I=0,0.3$ and $0.5$.} \label{mmval}
\end{table} 

In \cref{mmval} one can observe that, at T = 0 MeV and for given density of the medium, the magnitude of effective magnetic moments increase with rise of asymmetry parameter upto $I=0.3$ 
with exception of $\mu_{\Sigma^+}^*$ and $\mu_{\Sigma^0}^*$. This is because the system with larger asymmetry ($I=0.3$) becomes unbounded at low temperatures \cite{lopez}, leading to large increase in magnitude of effective magnetic moment of baryons as compared to the case of symmetric nuclear matter. The exceptional behavior of $\mu_{\Sigma^+}^*$ and $\mu_{\Sigma^0}^*$ is due to different behavior of their spin and orbital angular momentum wave functions used to calculate effective magnetic moments in the asymmetric matter as compared to behavior of wave functions of other baryons. At $\rho_B=\rho_0$, for increase in isospin asymmetry from $I=0.3$ to $I=0.5$  
the magnetic moments of $p,\Sigma^+$ and $\Xi^0$ show a decrease, whereas, the magnetic moments of $n,\Sigma^-,\Sigma^0$, $\Lambda$ and $\Xi^-$ show increase in magnitude, whereas, at $\rho_B=4\rho_0$, the magnitude of magnetic moments of $n,\Sigma^-,\Lambda$ show opposite behavior. This is due to different behaviors of their valence, sea and orbital angular momentum contributions at $4\rho_0$ as compared to their behavior at $\rho_0$. For instance, in case of neutron, at  $\rho_B=\rho_0$, the orbital angular momentum contribution causes the increase in magnitude of its magnetic moment, whereas, at $4\rho_0$ this contribution becomes so small that its increase do not come over the decreasing magnitudes of valence and sea quark contributions leading to decrease in magnitude of magnetic moment of neutron at $4\rho_0$. 
Further, at given finite density of the medium, the effect of rise of asymmetry parameter 
on effective magnetic moments becomes very small 
at T = 100 MeV (as the effective magnetic moments remain almost same at $I=0.3$ and $I=0.5$ especially at $\rho_B=4\rho_0$, as depicted by \cref{mmvaldifft}). This is due to the fact that the rise of temperature 
makes the nuclear matter less sensitive to the effect of density and isospin asymmetry of the medium \cite{lopez}.

In order to understand the effect of isospin asymmetry of the medium on sea quark orbital angular momentum, in \cref{ommt1,ommt2} we have plotted $\mu^{*}_{\rm orbital}$ as a function of baryon density for isospin asymmetry parameters $I=0, 0.3$ and $0.5$ and temperature T = 0. In case of nucleons, the magnitude of $\mu^{*}_{\rm orbital}$ is observed to be more for isospin asymmetry parameter $I$ = 0.5 as compared to $I$ = 0.
For $\Xi^-$ and $\Xi^0$ baryons, the magnitude of $\mu^{*}_{\rm orbital}$, for densities below $0.8\rho_0$, is observed to decrease as we move from symmetric to asymmetric nuclear matter, whereas above this density the trend becomes opposite. 
In case of $\Xi^0$ and $\Xi^-$ baryons, $\mu^{*}_{\rm orbital}$ has almost same magnitude at a given density, temperature and isospin asymmetry parameter. This is because of the fact that
 $\mu^*_{\rm orbital}$  is calculated through the product of valence quark spin polarizations and corresponding  orbital moments of quarks composing respective baryon.  The product of valence quark spin polarizations and corresponding  orbital moments of two $s$ quarks dominate over light $u$ and $d$ quarks in  $\Xi^-$ and $\Xi^0$ baryons. This  leads to a similar variation of $\mu^*_{\rm orbital}(\Xi^0)$ and   $\mu^*_{\rm orbital}(\Xi^-)$ as a function of density of the medium.
As can be seen from \cref{ommt2}, in case of $\Sigma$ and $\Lambda$ baryons, the magnitude of $\mu^{*}_{\rm orbital}$  in isospin asymmetric matter is very different as compared to symmetric nuclear matter.
  The main reason behind the observed behavior of $\mu^{*}_{\rm orbital}$ lies in the fact that the symmetry breaking parameter $\varepsilon$ has different values in symmetric and asymmetric nuclear matter. This leads to different values of quark spin polarizations in case of different octet baryon members and hence leads to different sea quark orbital angular momentum contributions.
    
Our calculations show that at temperature T = 0 MeV, for the rise of density of nuclear medium from $\rho_B=0$ to saturation density, effective magnetic moment of proton increases by $20\%$ in case of symmetric as well as asymmetric nuclear matter, which is consistent with cloudy bag model calculations showing the enhancement of proton magnetic moment with the rise of nuclear matter density in the range of $2-20\%$ \cite{ryu3}. On comparing with our previous calculations for symmetric nuclear matter \cite{happy}, the enhancement of magnetic moment is less in the present work. This decrease is due to inclusion of pion loop effect in the magnetic moment of baryons. 
At T = 0 MeV, for rise of density from $\rho_B=0$ to $\rho_0$, the percentage increase in  
the magnitude of effective magnetic moment of $\Sigma$, $\Xi$ and $\Lambda$ are $12\%$, $11\%$, $17\%$, respectively. 
This behavior is completely different from that in case of QMC calculations, where the magnitude of $\mu_{\Lambda}^*$ decreases by $0.7\%$. However, in case of modified QMC calculations the magnitude of magnetic moment increases by $10\%$ \cite{ryu3} 
which may be due to the model dependence of effective quark masses.
%
 \begin{table}
  \begin{tabular}{|c|c|c|c|c|c|c|c|c|}  
 \hline 
  & {$\rho_B=0$} & \multicolumn{3}{c|}{$\rho_B=\rho_0$} & \multicolumn{3}{c|}{$\rho_B=4\rho_0$}    \\ 
 \hline
  & $I=0$ & $I=0$ & $I=0.3$ & $I=0.5$ & $I=0$ & $I=0.3$ & $I=0.5$  \\ 
 \hline 
 $\mu_p^*(\mu_N)$  & $2.55$ & $3.01$ & $3.02$ & $3.02$ & $3.43$ & $3.49$ & $3.47$   \\ 
 \hline 
 $\mu_n^*(\mu_N)$  & $-1.93$ & $-2.19$ & $-2.22$ & $-2.23$ & $-2.45$ & $-2.52$ & $-2.52$   \\ 
 \hline 
 $\mu_{\Sigma^+}^*(\mu_N)$  & $2.60$ & $2.88$ & $2.87$ & $2.86$ & $3.08$ & $3.05$ & $3.04$    \\ 
 \hline 
 $\mu_{\Sigma^-}^*(\mu_N)$  & $-1.26$ & $-1.28$ & $-1.30$ & $-1.31$ & $-1.23$ & $-1.32$ & $-1.32$   \\ 
 \hline 
 $\mu_{\Sigma^0}^*(\mu_N)$  & $-1.43$ & $-1.68$ & $-1.67$ & $-1.67$ & $-1.85$ & $-1.74$ & $-1.74$   \\ 
 \hline 
 $\mu_{\Xi^0}^*(\mu_N)$  & $-1.32$  & $-1.44$ & $-1.46$ & $-1.46$ & $-1.55$ & $-1.61$ & $-1.60$  \\ 
 \hline 
 $\mu_{\Xi^-}^*(\mu_N)$  & $-0.53$ & $-0.57$ & $-0.58$ & $-0.58$ & $-0.60$ & $-0.62$ & $-0.62$   \\ 
 \hline 
 $\mu_{\Lambda}^*(\mu_N)$  & $-0.55$ & $-0.63$ & $-0.64$ & $-0.65$ & $-0.71$ & $-0.77$ & $-0.77$   \\ 
 \hline   
 \end{tabular} 
\caption{Effective magnetic moments of octet baryons at T = 100 MeV and $\rho_B=0, \rho_0$ and $4\rho_0$, for asymmetry parameters $I=0,0.3$ and $0.5$.} \label{mmvaldifft}
\end{table} 
At given finite density and asymmetry parameter, the magnitude of effective magnetic moments of baryons decrease as a function of temperature.
This is evident from comparison of the values of effective magnetic moments at $\rho_B=\rho_0$ and $4\rho_0$, as given in \cref{mmval} and \cref{mmvaldifft}. For example, in symmetric nuclear matter at nuclear saturation density, the magnitude of effective magnetic moment of nucleons decreases by $2.5\%$ as the temperature increases from zero to 100 MeV. In case of $\Sigma^{\pm}$  baryons, at $\rho_B=\rho_0$, the magnitude of effective magnetic moment decreases by $1.5\%$ as the temperature increases from T = 0 MeV to T = 100 MeV. Further, in case of  $\Xi^0$ and $\Xi^-$ this decrease is only $1.6\%$. 
On the other hand, at $\rho_B=4\rho_0$ and for given asymmetry parameter, the change in effective value of magnetic moment of baryons is very small as a function of temperature as compared to the case of $\rho_B=\rho_0$. This can be understood as follows. 

At $\rho_B=0$, the effective quark mass remains almost same with the rise of temperature upto certain value of temperature, because the thermal distribution functions alone effect the self energies of constituent quarks and hence decreasing the effective quark masses (increasing the effective magnetic moment of baryons). However, with the rise of density, another contribution starts coming from higher momentum states due to which the effective magnetic moments decrease (as the effective masses of quarks increase) \cite{AMU}. Thus, due to these two contributions, i.e., thermal distribution function and higher momentum states, the behavior of effective magnetic moment of baryons is reversed with the rise of temperature at finite value of density of medium as compared to its behavior at zero baryonic density. Further, for still high densities, i.e., $4\rho_0$ or more, the variation of effective magnetic moment of baryons become insensitive to the variation in effective mass of constituent quarks. This can be due to second order chiral phase transition at higher densities and temperatures. If one considers the low density and higher temperature regime near the critical temperature, the phase transition is of first order. However, in the higher baryonic density regime for densities above $4\rho_0$ the phase transition becomes a second order chiral phase transition \cite{phase,phase1}. The effective masses of constituent quark appear to be very close to the chiral limit in this regime, and hence, show very small variation as a function of density or temperature of medium. This results in negligible variation in magnitude of magnetic moment with the variation of constituent quark masses. This observation is further justified by those expected in Ref. \cite{ryu1}, where medium modified baryonic magnetic moments using modified quark meson coupling model were calculated.

\section{Summary} \label{summ}
We have studied the masses and magnetic moments of octet baryons at finite density and temperature in an asymmetric nuclear matter by using the chiral SU(3) quark mean field approach. To obtain the vacuum values of magnetic moments of baryons comparable to experimental data, we have calculated the magnetic moments of baryons which further include the contribution from valence quarks, sea quarks and orbital angular momentum of sea quarks. 

With the introduction of isospin asymmetry in the nuclear medium, the effective masses of  $u$ quark and $d$ quark show a splitting which rises with the rise of isospin asymmetry of the medium. This splitting of effective quark masses also leads to splitting of effective baryon masses in  isospin multiplets in the baryon octet. We find that the coupling between quarks and mesons plays an important role in understanding the behavior of strongly interacting matter.  The present approach is based on mean field approximation, in which we consider constant coupling between quarks and mesons. It will be interesting to study the behavior of dense matter beyond mean field approximation. It is argued in literature that in the presence of magnetic field the coupling constants decrease with the strength of magnetic field (through thermo-magnetic corrections to couplings at one loop level), leading to decrease in dynamical (effective) mass of constituent quarks, especially at high temperatures. This can lead to decrease in critical temperature and hence the phenomenon of inverse magnetic catalysis \cite{lsm,contact46,contact47,contact48,contact49,bali,mcimcqcd}.
The effective magnetic moments of baryons are found to increase with the increase of isospin asymmetry of the medium upto $I=0.3$ (with exceptions of $\mu_{\Sigma^+}^*$ and $\mu_{\Sigma^0}^*$) as compared to corresponding values in symmetric medium. For further rise in isospin asymmetry from $I=0.3$ to $I=0.5$, the magnetic moments of $n,\Sigma^-,\Lambda$ show different behaviors at $\rho_0$ and $4\rho_0$. 
This is due to attractive potential provided by $\delta$ meson field which makes pure neutron matter bounded like symmetric nuclear matter. Further, the increase of temperature decreases the effect of isospin asymmetry on the masses as well as magnetic moments. This is because of the reduced compressibility of nuclear matter at finite temperature as compared to that at zero temperature. 
 Furthermore, the variation of effective magnetic moments of baryons as a function of temperature is negligible for nuclear matter density higher than $4\rho_0$ indicating second order phase transition at higher densities \cite{wang neu}. The present approach along with Polyakov loop corrections (which becomes important near critical temperature) can be used to study the phenomena like chiral restoration, deconfinement phase transition and inverse magnetic catalysis \cite{recommend,hadrongas,imcpnjl,effective}. It will be interesting to study the behavior of effective magnetic moment of baryons obtained in the present work, in the presence of magnetized neutron star matter as modification of baryon magnetic moment can lead to changes in neutron star mass as well as radius \cite{ryu1}. 



\begin{thebibliography}{1}

\section*{References}
\bibitem{bordbar} 
G. H. Bordbar, H. Nadgaram , Research in Astron. Astrophys. {\bf 12}, 345 (2012).

\bibitem{horst} 
H. Muller, B. D. Serot, Phys. Rev. C {\bf 52}, 2072 (1995).

\bibitem{friedman} 
B. Friedman, V. R. Pandharipande, Nucl. Phys. A {\bf 361}, 502 (1981).

 \bibitem{ryu1}
 C. Y. Ryu and K. S. Kim, Phys. Rev. C {\bf 82}, 025804 (2010).
 
\bibitem{mcimcqcd} 
N. Mueller, J. M. Pawlowski, Phys. Rev. D {\bf 91}, 116010 (2015). 

\bibitem{friese} 
V. Friese, Nucl. Phys. A {\bf 774}, 377 (2005).

\bibitem{zhan} 
W. Zhan \textit{et al.}, Int. J. Mod. Phys. E {\bf 15}, 1941 (2006).

\bibitem{yano} 
Y. Yano , Nucl. Instrum. Methods B {\bf 261}, 1009 (2007).





\bibitem{rahul} 
A. Kumar, R. Chhabra, Phys. Rev. C {\bf 92}, 035208 (2015).

\bibitem{eng} 
L. ENGVIK \textit{et al.}  The Astrophysical Journal {\bf 469}, 794 (1996).

\bibitem{dave} 
 D. Davesne, A. Pastore, J. Navarro, A$\&$A {\bf 585}, A83 (2016) .
 
\bibitem{sahu} 
S. K. Sahu, Journal of Modern Physics {\bf 6}, 1350 (1995).  

\bibitem{lopez} 
J. A. Lopez \textit{et al.} Phys. Rev. C {\bf 89}, 024611 (1995).

\bibitem{shlomo} 
S. Shlomo, Journal of Physics : Conference Series {\bf 337}, 012014 (2012).

\bibitem{fabio} 
F. L. Braghin, Brazilian Journal of Physics {\bf 33(2)}, 255 (2003).

\bibitem{lenske} 
A. Fedoseen, H. Lenske, Phys. Rev. C {\bf 91}, 034307 (2015).

\bibitem{jeremy} 
C. Wellenhofer, J. W. Holt, N. Kaiser, Phys. Rev. C {\bf 92}, 1, 015801 (2015).


\bibitem{savage} 
M. J. Savage, Nucl. Phys. A {\bf 700}, 359 (2002).

\bibitem{ss}
S. Sahu, Revista Mxicana De Fisica {\bf 48}, 48 (2002).

\bibitem{felix}
F. Schlumpf, Phys. Rev. D {\bf 48} , 4478 (1993).

\bibitem{wrb}
W. R. B. de Araujo \textit{et al.}, Brazilian Journal of Physics {\bf 34} , 871  (2004).

\bibitem{hack}
E. J. Hackett-Jones, D. B. Leinweber, A. W. Thomas, Phys. Lett. B {\bf 489}, 143 (2000).

\bibitem{jg}
J. G. Contreras, R. Huerta, Revista Mxicana De Fisica {\bf 50} , 490 (2004).

\bibitem{jun}
H. E. Jun, Dong Yu-Bing, Commun. Theor. Phys. {\bf 43} , 139 (2005).

\bibitem{lks}
L. K. Sharma, C. Mai, J. Sci {\bf 34} , 13 (2007).

\bibitem{mulders}
P. J. Mulders, Phys. Reports  {\bf 185}, 83 (1990).

\bibitem{lawley} 
S. Lawley, W. Bentz, A. W. Thomas, Nucl. Phys. Proc. Suppl. {\bf 141}, 29 (2005).

\bibitem{guic1} 
P. A. M. Guichon, Phys. Lett. B {\bf 200}, 235 (1988).

\bibitem{guic2} 
P. A. M. Guichon, K. Saito, E. N. Rodionov, A. W. Thomas, Nucl. Phys. A {\bf 601}, 349 (1996).

\bibitem{saito} 
K. Saito, A. W. Thomas, Phys. Rev. C {\bf 57}, 2757 (1995).

\bibitem{panda} 
P. K. Panda \textit{et al.}, Phys. Rev. C {\bf 68}, 015201 (2003).


\bibitem{smith} 
J. R. Smith, G. A. Miller, Phys. Rev. C {\bf 70}, 065205 (2004).

\bibitem{mil}
G. A. Miller, Phys. Rev. C {\bf 66}, 032201 (2002).

\bibitem{bub}
M. Buballa, Phys. Rept. {\bf 407}, 205 (2005).

\bibitem{serot2} 
R. J. Furnstanl, B. D. Serot, Phys. Rev. C {\bf 41}, 262 (1990).

\bibitem{bodmer} 
A. R. Bodmer, Nucl. Phys. Rev. A {\bf 547}, 447 (1992).

\bibitem{wang}
W. Ping, Z. Z. Ye and Y. Y. Wen, Commun. Theor. Phys. {\bf 36}, 71 (2001).

\bibitem{wang2}
P. Wang \textit{et al.}, Phys. Rev. C {\bf 70}, 015202 (2004).


 
  \bibitem{ryu2}
 C. Y. Ryu, M. K. Cheoun, C. H. Hyun, Journal of Korean Physical Society {\bf 54}, 1448 (2009).
 
\bibitem{happy}
H. Singh, A. Kumar, H. Dahiya, Chinese Physics C, {\bf 41}, 094104 (2017). 

\bibitem{chen} 
L. W. Chen, C. M. Ko, B. A. Li, Phys. Rev. C {\bf 80}, 014322 (2009).

\bibitem{typel} 
S. Typel, B. A. Brown, Phys. Rev. C {\bf 64}, 027302 (2001).

\bibitem{asywang} 
P. Wang \textit{et al.}, Nucl. Phys. A {\bf 748} 226 (2005).

\bibitem{har}
H. Dahiya, M. Gupta, Phys. Rev. D {\bf 64}, 014013 (2001).

\bibitem{cheng1} 
 T. P. Cheng, L. F. Li, Phys. Rev. Lett. {\bf 80}, 2789 (1998).


\bibitem{kubis} 
S. Kubis, M. Kutschera, Phys. Lett. B {\bf 399}, 191 (1997).

\bibitem{greco} 
V. Greco, M. Colonna, M. Di Toro, Phys. Rev. C {\bf 67}, 015203 (2003).

\bibitem{liu} 
Liu, B. Greco \textit{et al.}, Phys. Rev. C {\bf 65}, 335 (2002).

\bibitem{grigor1} 
G. B. Alaverdyan, Gravit. Cosmol. {\bf 15}, 5 (2009).

\bibitem{grigor2} 
G. Alaverdyan, Research in Astron. Astrophys. {\bf 10}, 1255 (2010).

\bibitem{qmcmm}
K. Saito, K. Tsushima, A. W. Thomas, Prog. Part. Nucl. Phys. {\bf 58},1 (2007).
 








\bibitem{aarti}
A. Gridhar, H. Dahiya, M. Randhawa, Phys.Rev. D {\bf 92} 3, 033012 (2015).

\bibitem{cheng}  
T. P. Cheng and Ling Fong Li, Phys. Rev. Lett. {\bf 74}, 2872 (1995).


\bibitem{cqm}
J. Linde, T. Ohlsson, H. Snellman, Phys. Rev. D {\bf 57}, 092009 (1998).

\bibitem{miyazaki}
\textit{$https://www.ma.utexas.edu/mp\_arc/c/05/05-216.pdf$}
  
  


 








\bibitem{song} 
H. Q. Song, R. K. Su, Phys. Lett. B {\bf 358}, 179 (1995).




\bibitem{rau}
P. Rau, \textit{et al}., J. Phys. G: Nucl. Part. Phys. {\bf 40}, 085001 (2013).





\bibitem{papag}
P. Papazoglou, D. Zschiesche, S. Schramm, J. Schaffner-Bielich,
	H. St\"ocker, and W. Greiner, Phys. Rev. C {\bf 59},  411  (1999).



\bibitem{compress}
  J. Piekarewicz, M. Centelles, Phys. Rev. C {\bf 79}, 054311 (2009).

\bibitem{johan}
  J. Linde, T. Ohlsson and Hakan Snellman, Phys. Rev. D {\bf 57}, 452 (1998).

 \bibitem{ryu}
 C. Y. Ryu, C. H. Hyun, M. K. Cheoun, J. Phys. G {\bf 37}, 1052002 (2010).
 
 \bibitem{ryu3}
 C. Y. Ryu, C. H. Hyun, T. S. Park, S. W. Hong, Phys. Lett. B {\bf 674}, 122 (2009).



\bibitem{sw}
S. Weinberg, Physica A {\bf 96}, 327 (1979).

\bibitem{am}
A. Manohar, H. Georgi, Nucl. Phys. B  {\bf 234}, 189 (1984).


\bibitem{wang3}
P. Wang \textit{et al.} Nucl. Phys. A {\bf 688} 791 (2001).



\bibitem{chengsu3}
  T. P. Cheng, L. F. Li,  Phys. Rev. D {\bf 57}, 344 (1998).


\bibitem{mg} 
M. Gupta, N. Kaur, Phys. Rev. D {\bf 28}, 534 (1983).


\bibitem{abu}
M. Abu-Shady, A. K. Abu-Nab, American Journal of Physics and App. {\bf 46}, 1 (2014).

\bibitem{nldel}
T. Gaitanos, M. Kaskulov, Nucl. Phys. A {\bf 878}, 49 (2012).

\bibitem{dd}
W. H. Long, N. V. Giai, J. Meng, Phys. Lett. B {\bf 640}, 150 (2006).

\bibitem{qmcasy}
A. M. Santos, C. A. Providencia, P. K. Panda, Phys. Rev. C {\bf 79}, 045805 (2009).

\bibitem{compress}
S. Gandolfi, J. Carlson, S. Reddy, A. W. Steiner, R. B. Wiringa, Eur. Phys. J. A {\bf 50 2}, 10 (2014).

\bibitem{qmcasy22}
C. Fuchs and H. H. Wolter, Eur. Phys. J. A {\bf 30}, 5 (2006).

\bibitem{qmcasy23}
W. Zuo, A. Lejeune, U. Lombardo, J. F. Mathiot, Nucl. Phys.
A {\bf 706}, 418 (2002); T. Frick, Kh. Gad, H.Muther, P. Czerski,
Phys. Rev. C {\bf 65}, 034321 (2002).

\bibitem{mass41}
B. Y. Sun, W. H. Long, J. Meng, U. Lombardo, Phys. Rev. C {\bf 78}, 065805 (2008).

\bibitem{effmass}
B. Liu, V. Greco, V. Baran, M. Colonna, M. Di Toro, Phys. Rev. C {\bf 65}, 045201 (2002).


\bibitem{thomas}
 S. Theberge and A. W. Thomas, Phys. Rev. D {\bf 25}, 284 (1982); 
J. Cohen and H. J. Weber, Phys. Lett. B {\bf 165}, 229 (1985).
\bibitem{nobo}
 I. S. Sogami, N. Oh'yamaguchi, Phys. Rev. Lett.
{\bf 54}, 2295 (1985); K. T. Chao, Phys. Rev. D {\bf 41}, 920 (1990).

\bibitem{ld} 
L. Durand and P. Ha, Phys. Rev. D {\bf 58}, 013010 (1998).


 \bibitem{barik1}
 N. Barik, B. K. Dash, Phys. Rev. D {\bf 31}, 7 (1985).

 \bibitem{barik2}
 N. Barik \textit{et al.}, Phys. Rev. C {\bf 88}, 015206 (2013).

 \bibitem{wang neu}
 P. Wang \textit{et al.}, Phys. Rev. C {\bf 72}, 045801 (2005).
 	
\bibitem{AMU}
A. Mishra \textit{et al.} Eur. Phys, J. A {\bf 41}, 205  (2009).

\bibitem{nmc}
New Muon Collaboration, P. Amaudruz \textit{et al.}, Phys. Rev. Lett. {\bf 66}, 2712 (1991).
\bibitem{ling} 
L. F. Li and T. P. Cheng, Lecture Notes in Physics {\bf 512} (1998).
\bibitem{experiment} 
W-M Yao \textit{et al.}, Particle Data Group. J. Phys. G {\bf 33}, 1 (2006).
\bibitem{aarti}
A. Gridhar, H. Dahiya, M. Randhawa, Phys. Rev. D {\bf 92} 3, 033012 (2015).
\bibitem{oneloop}
G. Ramalho, K. Tsushima, Phys. Rev. D {\bf 84}, 054014 (2011).
\bibitem{har2}
Ikuo S. Sogami, N. Oh'yamaguchi, Phys. Rev. Lett {\bf 54}, 2295 (1985); Kuang-Ta Chao, Phys. Rev. D {\bf 41}, 920 (1990).
\bibitem{gupta}
 	M. Gupta, J. Phys. G {\bf 16}, L 213 (1990).
	
\bibitem{lsm}
A. Ayala, M. Loewe, R. Zamora Chinese J. Phys.: Conf. Ser. {\bf 720}, 012026 (2016).
\bibitem{contact46}
G. S. Bali \textit{et al.}, Phys. Rev. D {\bf 86}, 071502 (2012).
G. S. Bali \textit{et al.}, J. High Energy Phys. {\bf 02}, 044 (2012). 
\bibitem{contact47}
 R. L. S. Farias, K. P. Krein, M. B. Pinto, Phys. Rev. C {\bf 90}, 025203 (2014).
\bibitem{contact48}
A. Ayala, M. Loewe, R. Zamora, Phys. Rev. D {\bf 91}, 016002 (2014).
 A. Ayala, M. Loewe, R. Zamora, Phys. Rev. D {\bf 90}, 036001 (2014).
\bibitem{contact49}
M. Ferreira, P. Costa, O. Lourenço, T. Frederico, C. Providência, Phys. Rev. D {\bf 89}, 116011 (2014).
\bibitem{bali}
G. S. Bali \textit{et al.}, J. High Energy Phys. {\bf 1408}, 177 (2014).
\bibitem{recommend} 
A. N. Tawfik \textit{et al.}, Advances in High Energy Physics {\bf 2016}, 1381479 (2016).
\bibitem{hadrongas} 
G. Endrodi, J. High Energy Phys. {\bf 2013}, 23 (2013).
\bibitem{effective} 
J. Steinheimer \textit{et al.}, J. Phys. G: Nucl. Part. Phys. {\bf 38}, 035001 (2011).
\bibitem{imcpnjl} 
C. Providencia \textit{et al.}, Acta Phys. Polon. Supp. {\bf 8} 1, 207 (2015). 
\bibitem{phase}
J. Berger, D. U. Jungnickel, C. Wetterich, Eur. Phys. J. C {\bf 13}, 323 (2000).
\bibitem{phase1}
M. Abu-Shady, M. Soleiman, Physics of Particles and Nuclei Letters {\bf 10}, 683 (2013).	

 	 	 
\end{thebibliography}
\end{document}